\documentclass[twocolumn]{IEEEtran}
\usepackage{amsmath,amssymb,amsthm,epsfig,color,empheq,graphicx,graphics,balance,algorithm,algorithmic}
\usepackage{enumerate,url,wasysym,epstopdf,enumitem,array}
\usepackage{xcolor}
\usepackage{amsfonts}
\usepackage{adjustbox}
\usepackage{multirow}
\usepackage{accents}

\usepackage{float} 

\DeclareMathOperator{\diag}{dg}

\newtheorem{lemma}{Lemma}

\newcommand \bzero{\mathbf{0}}
\newcommand \bone{\mathbf{1}}
\newcommand \ba{\mathbf{a}}

\newcommand \bc{\mathbf{c}}

\newcommand \bef{\mathbf{f}} 
\newcommand \bg{\mathbf{g}}

\newcommand \bp{\mathbf{p}}
\newcommand \bq{\mathbf{q}}

\newcommand \bv{\mathbf{v}}

\newcommand \bx{\mathbf{x}}

\newcommand \bz{\mathbf{z}}
\newcommand \bA{\mathbf{A}}

\newcommand \bI{\mathbf{I}}

\newcommand \bR{\mathbf{R}}

\newcommand \bX{\mathbf{X}}

\newcommand \balpha{\boldsymbol{\alpha}}

\newcommand \bdelta{\boldsymbol{\delta}}

\newcommand \bsigma{\boldsymbol{\sigma}}

\newcommand \mcG{\mathcal{G}}

\newcommand \mcN{\mathcal{N}}

\newcommand \mcZ{\mathcal{Z}}

\newcommand \tbv{\tilde{\mathbf{v}}}

\newcommand \hbq{\hat{\mathbf{q}}}

\newcommand \bbq{\bar{\mathbf{q}}}

\newcommand \bbv{\bar{\mathbf{v}}}

\begin{document}

\title{Optimal Design of Volt/VAR Control Rules for Inverter-Interfaced Distributed Energy Resources}

\author{
    Ilgiz Murzakhanov,~\IEEEmembership{Member,~IEEE,}
    Sarthak Gupta,~\IEEEmembership{Graduate Student Member,~IEEE,}\\
    Spyros~Chatzivasileiadis,~\IEEEmembership{Senior Member,~IEEE,} and
    Vassilis Kekatos,~\IEEEmembership{Senior Member,~IEEE}

\thanks{Manuscript received October 22, 2022; and revised February 10, and April 23, 2023; accepted May 20, 2023. This work was supported in part by the ID-EDGe project, funded by Innovation Fund Denmark, Grant Agreement No. 8127-00017B, and the US National Science Foundation grant 2034137. I.~Murzakhanov and S.~Chatzivasileiadis are with the Department of Wind and Energy Systems, Technical University of Denmark. E-mails: \{ilgmu, spchatz\}@dtu.dk. S.~Gupta and V.~Kekatos are with the Bradley Dept. of ECE, Virginia Tech, Blacksburg, VA 24061, USA. E-mails: \{gsarthak,kekatos\}@vt.edu}
\thanks{Color versions of one or more of the figures is this paper are available online at {http://ieeexplore.ieee.org}.}
\thanks{Digital Object Identifier XXXXXX}
}	


\maketitle


\begin{abstract}
The IEEE 1547 Standard for the interconnection of distributed energy resources (DERs) to distribution grids provisions that smart inverters could be implementing Volt/VAR control rules among other options. Such rules enable DERs to respond autonomously in response to time-varying grid loading conditions. The rules comprise affine droop control augmented with a deadband and saturation regions. Nonetheless, selecting the shape of these rules is not an obvious task, and the default options may not be optimal or dynamically stable. To this end, this work develops a novel methodology for customizing Volt/VAR rules on a per-bus basis for a single-phase feeder. The rules are adjusted by the utility every few hours depending on anticipated demand and solar scenarios. Using a projected gradient descent-based algorithm, rules are designed to improve the feeder’s voltage profile, comply with IEEE 1547 constraints, and guarantee stability of the underlying nonlinear grid dynamics. The stability region is inner approximated by a polytope and the rules are judiciously parameterized so their feasible set is convex. Numerical tests using real-world data on the IEEE 141-bus feeder corroborate the scalability of the methodology and explore the trade-offs of Volt/VAR control with alternatives.
\end{abstract}




\section{Introduction}
\allowdisplaybreaks
Motivated by climate change concerns and rising fossil fuel prices, countries around the globe are integrating large amounts of solar photovoltaics and other distributed energy resources (DERs) into the grid. Unfortunately, the uncertain nature of photovoltaics and DERs can result in undesirable voltage fluctuations in distribution feeders. Inverters equipped with advanced power electronics can provide effective voltage regulation through reactive power compensation if properly orchestrated. This work aims at designing the Volt/VAR control rules for inverters, as recommended by the IEEE 1547.8 Standard~\cite{IEEE1547.8}, on a quasi-static basis to ensure their dynamic stability and real-time voltage regulation performance. 

Inverter-based voltage regulation has been extensively studied and adopted approaches can be classified as \emph{centralized}, \emph{distributed}, and \emph{localized}. \emph{Centralized} approaches entail communicating instantaneous load/solar data to the utility, solving an optimal power flow (OPF) problem to obtain optimal setpoints~\cite{FCL}, and communicating back to inverters. 
Although centralized approaches are able to compute optimal setpoints, they may incur high computation and communication overhead in real time. 
\emph{Distributed} approaches partially address these concerns by sharing the computational burden across inverters~\cite{Bolognani13,7963364}. However, they may need a large number of iterations to converge, which leads to delays in obtaining setpoints. Real-time OPF schemes where setpoints are updated dynamically have been shown to be effective~\cite{OPFpursuit,RTOPF} by computing fast an approximate solution, yet two-way communication is still necessary.


Controlling inverters using \emph{control rules} has been advocated as an effective means to reduce the computational overhead. In such a scheme, inverter setpoints are decided as a (non)-linear function of solar, load, and/or voltage data; see e.g., \cite{JKGD19}, \cite{GKJ2020} and references therein. Although such approaches reduce the computational burden, they still have high communication needs if driven by non-local data. To this end, there has been increased interest in local rules, i.e., policies driven by purely local data~\cite{Turitsyn11}. Perhaps not surprisingly, local control rules lack global optimality guarantees as established in~\cite{Guido16},~\cite{8667359}, yet they offer autonomous inverter operation.

As a predominant example of local control rules, the IEEE 1547.8 standard provisions that inverter setpoints can be selected upon Volt/VAR, Watt/VAR, or Volt/Watt rules~\cite{IEEE1547.8}. The recommended rules take a parametric, non-increasing, piecewise affine shape, equipped with saturation regions and a deadband. Albeit easy to implement, designing the exact shape of control curves is not an obvious task. Among the different control options, Volt/VAR rules could be considered most effective as voltage is the quantity to be controlled and also carries non-local information. Watt/VAR curves have been optimally designed before in~\cite{Jabr18,SKL19GM}. {The resulting optimization models involve products between continuous and binary variables, which can be handled exactly using McCormick relaxation (big-M trick) as in~\cite{SKL19GM}. On the other hand, designing Volt/VAR curves is more challenging as they incur a closed-loop dynamical system, whose stability needs to be enforced. Moreover, designing Volt/VAR curves gives rise to optimization models involving products between continuous variables, which are harder to deal with.}

Although Volt/VAR rules have been shown to be stable under appropriate conditions, their equilibria may not be optimal in terms of voltage regulation performance~\cite{XFLCL21}, \cite{FCL13}, \cite{Pedram13}. This brings about the need for systematically designing Volt/VAR curves and customizing their shapes based on grid loading conditions on a per-bus basis. Volt/VAR dynamics exhibit an inherent trade-off between stability and voltage regulation. In view of this, several works have suggested augmenting Volt/VAR rules with a delay component so that the reactive power setpoint $q(t)$ at time $t$ depends on voltage $v(t)$ as well as the previous setpoint $q(t-1)$. These so-termed \emph{incremental rules} have been studied and designed in~\cite{LQD14,FZC15,7361761,VKZG16,8365842}. Here we focus on non-incremental Volt/VAR curves to be compliant with the IEEE 1547.8 Standard. Reference~\cite{Baker18} designs stable Volt/VAR curves to minimize the worst-case voltage excursions when loads and solar generation lie within a polyhedral uncertainty set. This design task can be approximated by a quadratic program; however, Volt/VAR rules are oversimplified as affine, ignoring their deadband and/or saturation regions. For example, {reference~\cite{9781808} simultaneously optimizes affine Volt/VAR and nonlinear (polynomial) Volt/Watt rules. The proposed optimization program incorporates stability and adopts a robust uncertainty set design.}
Reference~\cite{9796576} considers the detailed model of Volt/VAR curves and integrates them into a higher-level OPF formulation to properly capture the behavior of Volt/VAR-driven DERs; nevertheless, here curve parameters are assumed fixed and are not designed. 

This work considers the optimal design of Volt/VAR control rules in single-phase distribution grids. Using a dataset of grid loading scenarios anticipated for the next 2-hr period, the goal is to centrally and optimally design Volt/VAR control curves to attain a desirable voltage profile across a feeder. The contributions are on three fronts: \emph{i)} Develop a scalable projected gradient descent algorithm to find near-optimal Volt/VAR control curves. The computed curves are customized per inverter location, comply with the detailed form and constraints provisioned by the IEEE 1547 Standard, and ensure stable Volt/VAR dynamics (Section~\ref{sec:solution}); \emph{ii)} Provide a polytopic representation for the dynamic stability region of Volt/VAR control rules (Section~\ref{subsec:model:stability}); and \emph{iii)} Select a proper representation of the Volt/VAR rule parameters so that stability and the IEEE 1547-related constraints are expressed as a convex feasible set (Section~\ref{sec:design}). The proposed design scheme is evaluated using numerical tests using real-world load and solar generation data on the IEEE 141-bus feeder.

The paper is organized as follows. Section~\ref{sec:model} reviews an approximate feeder model, the IEEE 1547 Volt/VAR rules, their steady-state properties, and expands upon their stability. Section~\ref{sec:design} states the task of optimal rule design and selects a proper parameterization of the rules. Section~\ref{sec:solution} presents an iterative algorithm based on projected gradient descent to cope with the optimal rule design task. Numerical tests are reported in Section~\ref{sec:tests} and conclusions are drawn in Section~\ref{sec:conclusions}.

\emph{Notation:} Column vectors (matrices) are denoted by lower- (upper-) case letters. Operator $\diag(\bx)$ returns a diagonal matrix with $\bx$ on its diagonal. Symbol $(\cdot)^\top$ stands for transposition; and $\bI_{N}$ is the $N \times N$ identity matrix.

\section{Feeder and Control Rule Modeling}\label{sec:model}

\subsection{Feeder Modeling}\label{subsec:feeder}
Consider a single-phase radial distribution feeder with $N+1$ buses hosting a combination of inelastic loads and DERs. Buses are indexed by set $\mcN:=\{1,\dots,N\}$. Let $v_n$ denote the voltage magnitude at bus $n$, and $p_n + j q_n$ the complex power injected at bus $n\in\mcN$. Let vectors $(\bv,\bp,\bq)$ collect the aforesaid quantities across all buses. To express the dependence of voltages on power injections, we adopt the widely used linearized grid model~\cite{TJKT-SG21}
\begin{equation}\label{eq:ldf}
\bv\simeq \bR\bp+\bX\bq+v_0\bone    
\end{equation}
where $v_0$ is the substation voltage. Matrices $(\bR,\bX)$ are symmetric positive definite with positive entries. They depend on the feeder topology and line impedances, which are assumed fixed and known for the control period of interest. {Symbol $\bone$ denotes a vector of all ones and of appropriate length.}

The vectors of power injections can be decomposed as 
\begin{equation*}
    \bp = \bp^g - \bp^\ell\quad\quad \text{and}\quad \quad \bq = \bq^g - \bq^\ell
\end{equation*}
where $\bp^g+j\bq^g$ is the complex power injected by inverter-interfaced DERs, and $\bp^\ell+j\bq^\ell$ is the complex power consumed by uncontrollable loads.

Volt/VAR control amounts to adjusting $\bq^g$ with the goal of maintaining voltages around one per unit (pu) despite fluctuations in $(\bp^g,\bp^\ell,\bq^\ell)$. With this control objective in mind, let us rewrite \eqref{eq:ldf} as
\begin{equation}\label{eq:ldf2}
\bv= \bX\bq^g+\tbv= \bX\bq+\tbv
\end{equation}
where with a slight abuse in notation, we will henceforth denote $\bq^g$ simply by $\bq$. Moreover, vector $\tbv:=\bR(\bp^g - \bp^\ell)-\bX\bq^\ell+v_0\bone$ captures the effect of current grid loading conditions on voltages, and will be referred to simply as the vector of \emph{grid conditions}.

\subsection{Volt/VAR Rules}\label{subsec:rules}
The IEEE 1547 standard provisions DERs to provide reactive power support according to four possible modes: \emph{i)} constant reactive power; \emph{ii)} constant power factor; \emph{iii)} active power-dependent reactive power (watt-var); and \emph{iv)} voltage-dependent reactive power (volt-var) mode. Mode \emph{i)} is invariant to grid conditions. Modes \emph{ii)} and \emph{iii)} do adjust reactive injections, yet adjustments depend solely on the active injection of the individual DER. On the contrary, mode \emph{iv)} adjusts the reactive power injected by each DER based on its voltage magnitude. Albeit measured locally, voltage carries non-local grid information and constitutes the quantity of control interest anyway. Hence, our focus is on Volt/VAR control rules.

\begin{figure}[t]
	\centering
	\includegraphics[scale=0.3]{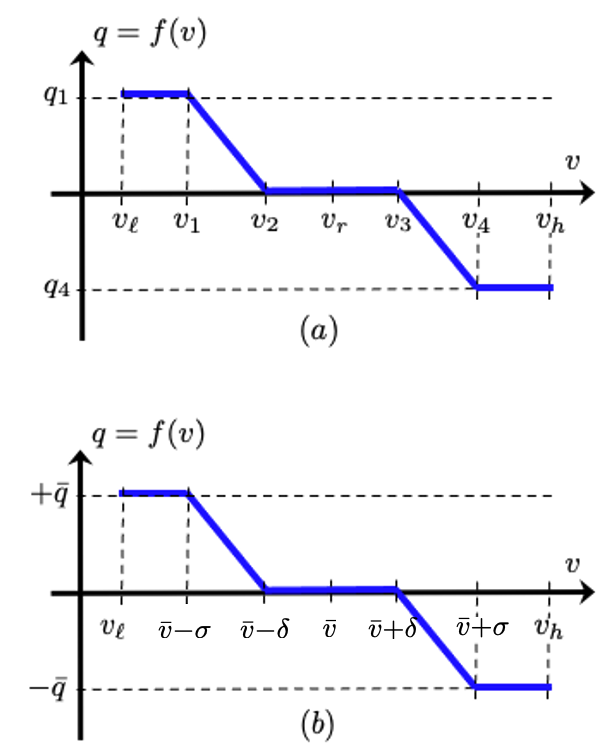}
	\caption{(a) IEEE 1547 Volt/VAR rule~\cite{IEEE1547.8}; and (b) its symmetric version.}
	\label{fig:curve}
\end{figure}

Per the IEEE 1547 standard~\cite{IEEE1547.8}, a Volt/VAR control rule is described by the piecewise affine function shown on Fig.~\ref{fig:curve}(a). This plot shows the dependence of reactive injection on local voltages over a given range of operating voltages $[v_\ell,v_h]$. The curve is described by voltage points $(v_1,v_2,v_r,v_3,v_4)$ and reactive power points $(q_1,q_4)$. To simplify the presentation, let us suppose the Volt/VAR curve is odd symmetric around the axis $v=\bar{v}=v_r$. The simplified rule shown in Fig.~\ref{fig:curve}(b) can be described by four parameters: the reference voltage $\bar{v}$; the deadband voltage $\bar{v}+\delta$; the saturation voltage $\bar{v}+\sigma$; and the saturation reactive power injection $\bar{q}$. Per the standard, the tuple $(\bar{v},\delta,\sigma,\bar{q})$ expressed in pu is constrained as
\begin{subequations}\label{eq:1547con2}
\begin{align}
0.95 &\leq \bar{v} \leq 1.05\\
0 &\leq   \delta     \leq 0.03\\
\delta+0.02  &\leq   \sigma     \leq 0.18\\
&0\leq \bar{q}\leq \hat{q}.\label{eq:1547con2:q}
\end{align}
\end{subequations}
Constraint~\eqref{eq:1547con2:q} ensures the extreme reactive setpoints are within the reactive power capability $\hat{q}$ of the DER. It is worth emphasizing here that the Volt/VAR curve can be designed to saturate at $\bar{q}$ that can be lower than $\hat{q}$. The standard also specifies default settings as $\delta=0.02$, $\sigma=0.08$, $\bar{v}=1$, and $\bar{q}=\hat{q}=0.44\bar{p}$, where $\bar{p}$ is the per-unit kW rating of the DER.

The rule segment over $[\bar{v}+\delta,\bar{v}+\sigma]$ can also be written as
\begin{equation}\label{eq:slope}
q=-\alpha(v-\bar{v}-\delta)\quad \text{where} \quad \alpha=\frac{\bar{q}}{\sigma-\delta}>0.
\end{equation}
The rule segment over $[\bar{v}-\sigma,\bar{v}-\delta]$ is $q=-\alpha(v-\bar{v}+\delta)$. We will see later that the slope $\alpha$ is crucial for stability. 

Although the standard offers flexibility in the design of Volt/VAR curves, it is not clear to utilities and software vendors how to optimally tune such control settings. The tuple $(\bar{v},\delta,\sigma,\bar{q})$ can be customized on a per bus basis. Let $q_n=f_n(v_n)$ denote the Volt/VAR rule for bus $n$. The rule for DER $n$ is  parameterized by $(\bar{v}_n,\delta_n,\sigma_n,\bar{q}_n)$. Let vectors $(\bbv,\bdelta,\bsigma,\bbq)$ collect the rule parameters across all buses. 

When Volt/VAR-controlled DERs interact with the electric grid, they give rise to the nonlinear dynamical system~\cite{XFLCL21}
\begin{subequations}\label{eq:dynamics1}
\begin{align}
&\bv^t = \bX\bq^t + \tbv\label{eq:dynamics1:a}\\
&\bq^{t+1} = \bef(\bv^t).\label{eq:dynamics1:b}
\end{align}
\end{subequations}
where the $n$-th entry of \eqref{eq:dynamics1:b} denotes the Volt/VAR rule $q_n=f_n(v_n)$ for DER $n$. Given the nonlinear dynamics of \eqref{eq:dynamics1}, three questions arise: \emph{q1)} Is the system stable? {In other words, if initiated at some $\bv^0$ for $t=0$, DERs run Volt/VAR rules $q_n=f_n(v_n)$ for all $n$, and grid experiences loading conditions $\tbv$, do the nonlinear dynamics of \eqref{eq:dynamics1} reach an equilibrium where $\bv^t=\bv_\text{eq}$ and $\bq^t=\bq_\text{eq}$ for all $t>T$ for some $T$? During the interval $t\in[0,T]$, the grid loading conditions $\tbv$ are assumed time-invariant assuming Volt/VAR dynamics are faster than load/solar variations.} \emph{q2)} If stable, what is its equilibrium? and \emph{q3)} Is that equilibrium useful for voltage regulation? These questions have been addressed in \cite{XFLCL21}, \cite{FCL13}. We review and build upon those answers.

\subsection{Stability of Volt/VAR Rules}\label{subsec:model:stability}
Regarding \emph{q1)}, let vector $\balpha$ collect all slope parameters $\alpha_n$ and define the diagonal matrix $\bA:=\diag(\balpha)$. The nonlinear dynamics in \eqref{eq:dynamics1} are stable if $\|\bA\bX\|_2<1$, where $\|\bA\bX\|_2$ is the maximum singular value of $\bA\bX$; see~\cite{XFLCL21,FCL13,Pedram13} for proofs of local and global exponential stability. If DERs are installed only on a subset $\mcG\subseteq\mcN$ of buses, the condition becomes $\|\bA\bX_{\mcG\mcG}\|_2<1$, where $\bX_{\mcG\mcG}$ is obtained from $\bX$ by keeping only its rows/columns associated with the buses in $\mcG$. To simplify the presentation, we will henceforth assume $\mcG=\mcN$, and elaborate when needed otherwise. 

Because it is hard to satisfy $\|\bA\bX\|_2<1$ as a strict inequality, one may want to tighten the constraint as~\cite{Baker18}
\begin{equation}\label{eq:stability1}
\|\bA\bX\|_2\leq 1-\epsilon
\end{equation}
for some positive $\epsilon$. Constraint~\eqref{eq:stability1} can be expressed as a linear matrix inequality (LMI) on $\balpha$ as
\begin{equation*}
\left[\begin{array}{cc}
    (1-\epsilon)\bI & \bA\bX \\
     \bX\bA & (1-\epsilon)\bI    
\end{array}\right]\succ \bzero.
\end{equation*}
This is because by Schur's complement, the LMI is equivalent to $(1-\epsilon)^2\bI\succeq \bX\bA^2\bX$, or equivalently, $\|\bA\bX\|_2=\sqrt{\lambda_\text{max}(\bX\bA^2\bX)}\leq 1-\epsilon$. To avoid the computational complexity of an LMI, reference~\cite{XFLCL21} surrogated the LMI for $\epsilon=0$ by the linear inequality $\|\balpha\|_{\infty}\cdot \|\bX\bone\|_{\infty}<1$. This inequality may be conservative as it upper bounds all slopes $\alpha_n$ by the same constant $\|\bX\bone\|_{\infty}^{-1}$. We next propose a tighter restriction of the LMI involving only linear inequality constraints.

\begin{lemma}\label{le:stability2}
The dynamical system in \eqref{eq:dynamics1} is stable if
\begin{subequations}\label{eq:stability2}
\begin{align}
&\bX\balpha \leq (1-\epsilon)\cdot \bone\label{eq:stability2:C1}\\
&\balpha \leq  (1-\epsilon)\cdot\left[\diag(\bX\bone)\right]^{-1}\bone\label{eq:stability2:C2}
\end{align}
\end{subequations}
with the inequalities understood entrywise.
\end{lemma}

\begin{IEEEproof}
Matrices $\bA$ and $\bX$ have non-negative entries. By a rendition of H\"{o}lder's inequality, it holds that
\[\|\bA\bX\|_2^2 \leq \|\bA\bX\|_1\cdot \|\bA\bX\|_\infty\]
where $\|\bA\bX\|_1$ is the maximum column-sum and $\|\bA\bX\|_{\infty}$ is the maximum row-sum of matrix $\bA\bX$. It is not hard to verify that \eqref{eq:stability2:C1} implies $\|\bA\bX\|_1\leq 1-\epsilon$, while \eqref{eq:stability2:C2} implies $\|\bA\bX\|_{\infty}\leq 1-\epsilon$, so \eqref{eq:stability1} follows.
\end{IEEEproof}

Note that previous works have suggested that \eqref{eq:stability2:C2} alone is sufficient to ensure stability~\cite{FCL13,8365842}. The next counterexample shows that \eqref{eq:stability2:C1} is needed as well. Suppose a toy feeder with three buses. Bus 1 is connected to the substation, and bus 2 is connected to bus 1. Both lines have 1-pu reactance. Matrix $\bX$ can be found to be
\[\bX=\left[\begin{array}{cc}
1 & 1\\
1 & 2
\end{array}\right].\]
The slope vector $\balpha_1=(1-\epsilon)[\tfrac{1}{2}~ \tfrac{1}{3}]^\top$ satisfies \eqref{eq:stability2:C2} with equality. Nonetheless, it does not satisfy \eqref{eq:stability1} as $\|\diag(\balpha_1)\bX\|_2= (1-\epsilon)\cdot 1.014\geq 1-\epsilon$. Indeed, vector $\balpha_1$ violates \eqref{eq:stability2:C1} as $\bX\balpha_1=(1-\epsilon)\cdot[0.833~1.1167]^\top$.



\subsection{Equilibrium Voltages and VAR Injections}
Regarding question \emph{q2)}, if stable, the dynamics of \eqref{eq:dynamics1} reach the equilibrium
\begin{subequations}\label{eq:equilibrium}
\begin{align}
\bv_\mathrm{eq}&=\bX\bq_\mathrm{eq} + \tbv\\
\bq_\mathrm{eq} &= \bef(\bv_\mathrm{eq}).
\end{align}
\end{subequations}
Interestingly, the equilibrium $\bq_\mathrm{eq}$ actually coincides with the unique minimizer of the convex quadratic program~\cite{FCL13,XFLCL21}
\begin{align}\label{eq:inner}
\bq_\mathrm{eq}=\arg\min_{\bq}~& V(\bq)+C(\bq)\\
\textrm{s.to}~&-\bbq\leq \bq \leq\bbq\nonumber
\end{align}
with the two components in the cost of \eqref{eq:inner} being defined as
\begin{subequations}\label{eq:10}
\begin{align}
V(\bq)&:=\frac{1}{2}\bq^\top \bX\bq +\bq^\top(\tbv-\bbv)\quad\text{and}\label{eq:Vn}\\
C(\bq)&:=\sum_{n\in\mcG}\frac{1}{2\alpha_n}q_n^2 +\delta_n|q_n|.\label{eq:Cn}
\end{align}
\end{subequations}

Let us elaborate on \eqref{eq:inner}: Under grid conditions $\tbv$ and if the DERs implement the Volt/VAR rules described by $(\bbv,\bdelta,\bbq,\balpha)$ under \eqref{eq:stability1}, the reached equilibrium $\bq_{\mathrm{eq}}$ is the $\bq$ that minimizes $V(\bq)+C(\bq)$ subject to $-\bbq\leq \bq\leq \bbq$. {Because $\bX$ is positive definite, the minimizer of \eqref{eq:inner} and thus the equilibrium $\bq_\mathrm{eq}$, are unique~\cite[Th.~4]{XFLCL21}.} \emph{Are DER injections $\bq_{\mathrm{eq}}$ and the associated equilibrium voltages $\bv_\mathrm{eq}=\bX\bq_{\mathrm{eq}}+\tbv$ desirable?} The fact that $\bq_{\mathrm{eq}}$ constitutes the minimizer of \eqref{eq:inner} provides useful intuition as suggested in~\cite{FCL13,7436387,7361761}. Component $C(\bq)$ in the objective penalizes excessive reactive power injections that can increase power losses. Component $V(\bq)$ on the other hand can be shown to be equal to~\cite{FCL13} 
\[V(\bq)=\frac{1}{2}(\bv-\bbv)^\top \bX^{-1}(\bv-\bbv) +\mathrm{constants}.\]
Because $\bX^{-1}\succ \bzero$, this is a \emph{rotated} $\ell_2$-norm of voltage deviations from reference voltages $\bbv$. For voltage regulation purposes, one would prefer minimizing $\|\bv-\bone\|_2^2$ rather than $V(\bq)$. Even if $V(\bq)$ was a reasonable proxy for voltage regulation, problem \eqref{eq:inner} includes also $C(\bq)$ in its objective. Then, to minimize $V(\bq)$, the component $C(\bq)$ should be diminished by sending $\bdelta$ to zero and $\balpha$ to infinity as commented in~\cite{FCL13,8365842}. That would cancel deadbands and cause stability issues as $\balpha$ is upper bounded by~\eqref{eq:stability1}.

When DERs are placed on a subset of buses, the objective component $V(\bq)$ in \eqref{eq:inner} should be altered as
\begin{align}\label{eq:Vn2}
V_\mcG(\bq)&:=\frac{1}{2}\bq^\top \bX_{\mcG\mcG}\bq+\bq^\top(\tbv_\mcG-\bbv)\nonumber\\
&=
\frac{1}{2}(\bv_\mcG-\bbv)^\top \bX_{\mcG\mcG}^{-1}(\bv_\mcG-\bbv) +\mathrm{constants}
\end{align}
where $(\bq_\mcG,\bv_\mcG,\tbv_\mcG)$ are the subvectors of $(\bq,\bv,\tbv)$ corresponding to buses with inverters comprising $\mcG$. Again, the objective component $V_\mcG(\bq)$ may not be good proxy for $\|\bv-\bone\|_2^2$.  
In a nutshell, the equilibrium $\bv_\mathrm{eq}$ reached by Volt/VAR rules may not be a desirable voltage profile. In essence, this is the price to be paid for allowing \emph{local} Volt/VAR rules, i.e., rules driven exclusively by local voltages rather than global information~\cite{FCL13,7436387}. Due to this locality, Volt/VAR rules can respond to solar/load fluctuations in real time without communicating with the operator. To improve the voltage profile, the operator can carefully design stable control rules so that the minimizer of \eqref{eq:inner} yields a more desirable voltage profile. Control rule parameters can be customized per bus and optimized centrally on quasi-static basis (e.g., every two hours). The optimal design of Volt/VAR curves should be solved centrally as the operator knows the feeder model and has estimates or historical data on the load/solar conditions to be experienced over the next hour. 

\section{Problem Formulation}\label{sec:design}
This section tackles optimal rule design in three steps: \emph{s1)} It first selects a convenient representation for the Volt/VAR rule parameters; \emph{s2)} It then defines the feasible set for these parameters; and \emph{s3)} Defines the cost to be optimized. 

Starting with step \emph{s1)}, let vector $\bz$ collect some parameterization of the Volt/VAR rules. The rule parameters $\bz$ should satisfy the IEEE 1547 and stability constraints of \eqref{eq:1547con2} and \eqref{eq:stability2}, respectively. All these constraints create the feasible set $\mcZ$ for $\bz$. We would like to parameterize $\bz$ in a way so that $\mcZ$ is convex. This would be useful later when we would like to project iterates of $\bz$ onto $\mcZ$. The rule for each DER has essentially four degrees of freedom. For example, the description in \eqref{eq:1547con2} sets $(\bbv,\bdelta,\bsigma,\bbq)$ as the free parameters, whereas problem \eqref{eq:inner} uses $(\bbv,\bdelta,\balpha,\bbq)$ instead. These parameterizations are equivalent as $(\delta_n,\sigma_n,\alpha_n,\bar{q}_n)$ are related via \eqref{eq:slope}. To end up with a convex $\mcZ$, we will use yet another third equivalent parameterization. Specifically, we introduce variable
\begin{equation}\label{eq:c}
c_n:=\frac{1}{\alpha_n}~\quad \forall n \in \mcG 
\end{equation}
per DER $n$. Let vector $\bc$ collect all $c_n$'s. The control rules can then be parameterized by
\begin{equation}\label{eq:z}
\bz:=(\bbv,\bdelta,\bsigma,\bc).
\end{equation}

Under the parameterization of \eqref{eq:z}, we proceed with step \emph{s2)} of defining feasible set $\mcZ$. If we translate the IEEE and stability constraints of \eqref{eq:1547con2} and \eqref{eq:stability2} over $\bz$, we can see that $\mcZ$ is described by constraints
\begin{subequations}\label{eq:set1}
\begin{align}
0.95\cdot \bone &\leq \bbv \leq 1.05\cdot\bone\label{eq:set1:first}\\
\bzero &\leq   \bdelta     \leq 0.03\cdot \bone\\
\bdelta+0.02\cdot \bone  &\leq   \bsigma     \leq 0.18\cdot \bone \label{eq:lb_sigmadelta}\\
\bsigma-\bdelta&\leq \diag(\hbq)\cdot\bc\label{eq:set1:q}\\
\bc&\geq \frac{1}{1-\epsilon}\bX\bone\label{eq:set1:low}\\
\sum_{m\in\mcG}X_{nm}\frac{1}{c_m}&\leq 1-\epsilon,\quad \forall n\in\mcG.\label{eq:set1:inverse}
\end{align}
\end{subequations}
Note constraint~\eqref{eq:set1:q} is equivalent to \eqref{eq:1547con2:q} since [cf.~\eqref{eq:slope}]
\[c_n=\frac{\sigma_n-\delta_n}{\bar{q}_n}>0.\]
{Moreover, constraint~\eqref{eq:set1:low} ensures $\bc$ is positive as matrix $\bX$ has positive entries.}

With the exception of \eqref{eq:set1:inverse}, the constraints in \eqref{eq:set1} are linear. Albeit convex, constraint \eqref{eq:set1:inverse} cannot be directly expressed in convex conic form. We introduce a set of auxiliary variables $a_n$ collected in vector $\ba$, and replace \eqref{eq:set1:inverse} by constraints
\begin{subequations}\label{eq:soc}
\begin{align}
&\bX\ba\leq (1-\epsilon)\cdot \bone\label{eq:soc:a}\\
&a_n c_n \geq 1,\quad \forall~n\in\mcG.\label{eq:soc:b}
\end{align}
\end{subequations}
Evidently, if $1/c_m\leq a_m$ for all $m$ from \eqref{eq:soc:b} and because $\bX$ has positive entries, we get that $\sum_{m\in\mcG}X_{nm}\frac{1}{c_m}\leq \sum_{m\in\mcG}X_{nm}a_m\leq 1-\epsilon$ for all $n$. Therefore, the constraints in \eqref{eq:soc} imply \eqref{eq:set1:inverse}. Constraint~\eqref{eq:soc:a} is linear, while \eqref{eq:soc:b} can be expressed as a second-order cone (SOC) constraint. Notice each $a_n$ may not necessarily be equal to $\alpha_n$; it only holds that $a_n\geq \alpha_n=1/c_n$. The feasible set $\mcZ_{\epsilon}$ is finally described as
\begin{equation}\label{eq:mcZ}
\mcZ_{\epsilon}:=\{\bz: (\bz,\ba)~\textrm{satisfying~\eqref{eq:set1:first}--\eqref{eq:set1:low} and \eqref{eq:soc}}\}
\end{equation}
where the subscript $\epsilon$ indicates the dependence of $\mcZ_{\epsilon}$ on $\epsilon$.

The stability condition in \eqref{eq:stability1} could have been handled as $\|\bA\bX\|_F\leq 1-\epsilon$ since $\|\bA\bX\|_2\leq \|\bA\bX\|_F$ as in \cite{Baker18}. Although $\|\bA\bX\|_F\leq 1-\epsilon$ can be imposed as a single convex quadratic constraint in $\balpha$, having $\balpha$ rather than $\bc$ in the parameterization would have rendered $\mcZ_{\epsilon}$ non-convex due to \eqref{eq:set1:q}. 

We continue with step \emph{s3)} to select a proper objective. To optimally select $\bz$, the operator samples $S$ scenarios of load/solar injections $(\bp_s^g,\bp_s^\ell,\bq_s^\ell)$ for $s=1,\ldots,S$, which are representative of the grid conditions to be experienced over the next two hours. Each tuple $(\bp_s^g,\bp_s^\ell,\bq_s^\ell)$ yields the non-controllable voltage term of \eqref{eq:ldf2}
\begin{equation}\label{eq:vtilde}
\tbv_s=\bR(\bp_s^g-\bp_s^\ell)-\bX\bq_s^\ell+v_0\bone.
\end{equation}
Given $\tbv_s$ and applying stable Volt/VAR control rules parameterized by $\bz$, the feeder reaches the equilibrium of \eqref{eq:equilibrium} for which we use the notation
\begin{subequations}\label{eq:equilibriums}
\begin{align}
\bv_s&=\bX\bq_s + \tbv_s\label{eq:equilibrium_v}\\
\bq_s &= \bef(\bv_s).\label{eq:equilibrium_q}
\end{align}
\end{subequations}

Rule parameters $\bz$ can then be selected as the minimizer of 
\begin{equation}\label{eq:outer}
\min_{\bz\in\mcZ_{\epsilon}}~F(\bz):=\frac{1}{2S} \sum_{s=1}^S \|\bv_s(\bz)-\bone\|_2^2.
\end{equation}
The objective sums up the squared voltage deviations from unity at equilibrium across all buses and scenarios. Parameter $\epsilon>0$ can be varied to find the desirable trade-off between voltage profile and stability considerations~\cite{Baker18,9781808}. Notation $\bv_s(\bz)$ captures the dependence of the equilibrium voltage $\bv_s$ for scenario $s$ on parameters $\bz$. Different from $V_\mcG(\bq)$ in \eqref{eq:Vn2}, problem \eqref{eq:outer} aims at minimizing voltage deviations across all buses, not only the buses equipped with inverters. Section~\ref{sec:solution} puts forth an iterative algorithm for solving \eqref{eq:outer}. {Before doing so, a discussion on how the proposed methodology can be extended to non-symmetric Volt/VAR curves is due.}

\color{black}
\subsection{Non-symmetric Volt/VAR Rules}\label{subsec:nonsymmetric}
It should be noted that the Volt/VAR rules provisioned by the IEEE Standard 1547 do not need to be odd symmetric around the reference voltage. We next sketch how the analysis and optimization models presented so far can be extended to accommodate non-symmetric Volt/VAR rules.

Non-symmetric rules can be described by a tuple of seven parameters per DER $n$ as $(\bar{v}_n,\delta_n^+,\alpha_n^+,\bar{q}_n^+,\delta_n^-,\alpha_n^-,\bar{q}_n^-)$. The superscripts $+$ and $-$ refer to the curve for $v_n>\bar{v}_n$ and $v_n<\bar{v}_n$, respectively. Similar to \eqref{eq:slope}, parameters $\sigma_n^+$ and $\sigma_n^-$ can be derived as 
\[\sigma_n^+=\frac{\bar{q}_n^+}{\alpha_n^+}+\delta_n^+\quad \text{and}\quad\sigma_n^-=\frac{\bar{q}_n^-}{\alpha_n^-}+\delta_n^-.\]
It is trivial to extend constraints \eqref{eq:set1:first}--\eqref{eq:set1:q} to the non-symmetric setting using the newly introduced variables. We expound on the non-trivial modifications.

For stability, according to~\cite{FCL13,XFLCL21}, condition $\|\bA\bX\|_2<1$ ensures non-symmetric Volt/VAR rules are stable if the $\alpha_n$'s on the diagonal of $\bA$ are defined as $\alpha_n:=\max\{\alpha_n^+,\alpha_n^-\}$. We next translate that result to the polytopic restriction of $\|\bA\bX\|_2<1-\epsilon$. Introduce $c_n^+=1/\alpha_n^+$ and $c_n^-=1/\alpha_n^-$, and collect them in vectors $\bc^+$ and $\bc^-$, accordingly. Stability can be enforced by: \emph{i)} duplicating \eqref{eq:set1:low} with $\bc$ being replaced by $\bc^+$ and $\bc^-$; and \emph{ii)} substituting \eqref{eq:soc:b} by 
\[a_n c_n^+\geq 1\quad \text{and}\quad a_n c_n^-\geq 1\quad \forall n.\]

Regarding the equilibrium of Volt/VAR dynamics with non-symmetric rules, it can be shown to be unique and provided as the minimizer of the convex program
\begin{align}\label{eq:innerpm}
\min_{\bq}~& V(\bq)+C_\mathrm{ns}(\bq)\\
\textrm{s.to}~&-\bbq^-\leq \bq \leq\bbq^+\nonumber
\end{align}
where $V(\bq)$ is exactly as in \eqref{eq:Vn}, while $C_\mathrm{ns}(\bq)$ is defined as
\[C_\mathrm{ns}(\bq):=\sum_{n\in\mcG}\frac{1}{2\alpha_n^+}\left[q_n\right]_{+}^2 +\delta_n^+\left[q_n\right]_{+}+\frac{1}{2\alpha_n^-}\left[q_n\right]_{-}^2 +\delta_n^-\left[q_n\right]_{-}\]
for the convex functions
\[\left[q_n\right]_{+}:=\max\{q_n,0\}\quad \text{and}\quad\left[q_n\right]_{-}:=\max\{-q_n,0\}.\]
The proof is similar to those in \cite{FCL13,XFLCL21,7436387}, and is therefore omitted.
\color{black}

\section{Solution Methodology}\label{sec:solution}
Problem~\eqref{eq:outer} is non-convex. In fact, it can be interpreted as a bilevel optimization problem as $\bv_s(\bz)=\bX\bq_s(\bz) + \tbv_s$ and $\bq_s(\bz)$ is the minimizer of the \emph{inner} problem in \eqref{eq:inner}. More specifically, problem \eqref{eq:outer} can be expressed as
\begin{align}\label{eq:bilevel}
\min_{\bz\in\mcZ_{\epsilon}}&~F(\bz):=\frac{1}{2S} \sum_{s=1}^S \|\bX\bq_s(\bz) + \tbv_s-\bone\|_2^2\\
\textrm{s.to}&~\textrm{$\bq_s(\bz)$ minimizes \eqref{eq:inner} for $\bv_s$ and $\bz$},~\forall s.\nonumber
\end{align}

{This bilevel program can be reformulated and solved as a mixed-integer program. To avoid the involved computational complexity, we resort to tackling the outer problem via gradient descent approach.} More precisely, to find a stationary point of \eqref{eq:bilevel}, we propose applying the projected gradient descent iterates
\begin{align}\label{eq:pgd}
\bz^{i+1}&:=\left[\bz^{i}-\mu\bg^i\right]_{\mcZ_{\epsilon}}
\end{align}
where $\mu>0$ is a step size; $\bg^i$ is the gradient of $F$ with respect to $\bz$ evaluated at $\bz^i$; and $[\cdot]_{\mcZ_{\epsilon}}$ denotes the projection operator onto set $\mcZ_{\epsilon}$. We next elaborate on the projection onto $\mcZ_{\epsilon}$, and on computing the gradient $\bg^i$.

The projection $[\bx]_{\mcZ_{\epsilon}}$ can be seen as an operator which takes a vector argument $\bx$ and returns the minimizer
\begin{equation}\label{eq:project}
[\bx]_{\mcZ_{\epsilon}}:=\arg\min_{\bz\in\mcZ_{\epsilon}}~\|\bx-\bz\|_2^2.
\end{equation}
Here $[\bx]_{\mcZ_{\epsilon}}$ is the projection of $\bx$ onto set $\mcZ_{\epsilon}$ defined in \eqref{eq:mcZ}. Note that for the iterations in \eqref{eq:pgd}, the argument of the projection would be $(\bz^{i}-\mu\bg^i)$ at iteration $i$. 

Since $\mcZ_{\epsilon}$ has been selected to be a convex set, the projection step in \eqref{eq:project} can be formulated as a second-order cone program (SOCP) per iteration $i$. We next compute the gradient $\bg^i$. The gradient of function $F$ defined in \eqref{eq:outer} with respect to $\bz$ is
\begin{equation}\label{eq:dF}
\nabla_{\bz} F= \frac{1}{S}\sum_{s=1}^S \left(\nabla_{\bz}\bv_s\right)^\top \left(\bv_s-\bone\right).
\end{equation}
The Jacobian $\nabla_{\bz}\bv_s$ can be computed via the chain rule as
\begin{equation}\label{eq:jac}
\nabla_{\bz}\bv_s=\bX\cdot\nabla_{\bz}\bq_s.
\end{equation}

Note that $\bq_s=\bef\left(\bv_s,\bz\right)$ depends both on voltages and Volt/VAR parameters. Moreover, voltages depend on $\bq_s$ through the equilibrium of the Volt/VAR dynamics of \eqref{eq:dynamics1}. Therefore, to compute the Jacobian matrix $\nabla_{\bz}\bq_s$, we need to compute total derivatives as
\begin{align*}
    \nabla_{\bz}\bq_s&=\frac{\partial\bef}{\partial \bv_s}\cdot\nabla_{\bz}\bv_s+\frac{\partial\bef}{\partial \bz}\cdot\nabla_{\bz}\bz\\
    &=\frac{\partial\bef}{\partial \bv_s}\cdot\bX\cdot\nabla_{\bz}\bq_s+\frac{\partial\bef}{\partial \bz}.
\end{align*}
The second step follows from \eqref{eq:jac} and $\nabla_{\bz}\bz=\bI$. Then, we get that
\begin{equation}\label{eq:DqDz}
\nabla_{\bz}\bq_s=\left(\bI-\frac{\partial\bef}{\partial \bv_s}\cdot\bX\right)^{-1}\cdot\frac{\partial\bef}{\partial \bz}
\end{equation}

Let us now compute the Jacobian matrices $\frac{\partial\bef}{\partial \bv_s}$ and $\frac{\partial\bef}{\partial \bz}$ appearing in the previous formula. The control rule $q_n=f_n(v_n)$ of DER $n$ depends on the Volt/VAR parameters and the voltage at this specific bus $n$. It does not depend on the Volt/VAR parameters or voltages at other buses. Therefore, to compute $\frac{\partial\bef}{\partial \bv_s}$ and $\frac{\partial\bef}{\partial \bz}$, it suffices to find
\[\frac{\partial f_n}{\partial v_n},~\frac{\partial f_n}{\partial \bar{v}_n},~\frac{\partial f_n}{\partial \delta_n},~\frac{\partial f_n}{\partial \sigma_n},~\frac{\partial f_n}{\partial c_n}\quad \text{for all}~n.\]
The remaining entries of $\frac{\partial\bef}{\partial \bv_s}$ and $\frac{\partial\bef}{\partial \bz}$ would be zero.

To compute the aforesaid partial derivatives, let us express the control rule in a more convenient form. Recall that if $u(v-v_0)$ is the Heaviside step function for voltage $v$ shifted by some parameter $v_0$, then $r(v-v_0):=(v-v_0)\cdot u(v-v_0)$ is a shifted ramp function. We also know that 
\[\frac{\partial r(v-v_0)}{\partial v}=u(v-v_0)~~\text{and}~~\frac{\partial r(v-v_0)}{\partial v_0}=-u(v-v_0).\]
These partial derivatives are undefined when $v=v_0$.

Building on the above, heed that the control rule can be expressed as the sum of four ramp functions
\begin{align*}
f_n(v_n)&=\frac{1}{c_n}\big[(r(v_n-\bar{v}_n-\sigma_n)-r(v_n-\bar{v}_n-\delta_n)\big]\\
&~+\frac{1}{c_n}\big[r(-v_n+\bar{v}_n-\delta_n)-r(-v_n+\bar{v}_n-\sigma_n)\big].
\end{align*}
It is then easy to compute the sought derivatives as
\begin{subequations}\label{eq:derivatives}
\begin{align}
\frac{\partial f_n}{\partial {v}_n}&=\frac{1}{c_n}\big[u(v_n-\bar{v}_n-\sigma_n)-u(v_n-\bar{v}_n-\delta_n)\big]\\
&+\frac{1}{c_n}\big[-u(-v_n+\bar{v}_n-\delta_n)+u(-v_n+\bar{v}_n-\sigma_n)\big] \notag\\
\frac{\partial f_n}{\partial \bar{v}_n}&=\frac{1}{c_n}\big[-u(v_n-\bar{v}_n-\sigma_n)+u(v_n-\bar{v}_n-\delta_n)\big]\\
&+\frac{1}{c_n}\big[u(-v_n+\bar{v}_n-\delta_n)-u(-v_n+\bar{v}_n-\sigma_n)\big] \notag \\
\frac{\partial f_n}{\partial \delta_n}&=\frac{1}{c_n}\big[u(v_n-\bar{v}_n-\delta_n)-u(-v_n+\bar{v}_n-\delta_n)\big]\\
\frac{\partial f_n}{\partial \sigma_n}&=\frac{1}{c_n}\big[u(-v_n+\bar{v}_n-\sigma_n)-u(v_n-\bar{v}_n-\sigma_n)\big]\\
\frac{\partial f_n}{\partial c_n}&=-\frac{f_n(v_n)}{c_n}=-\frac{q_n}{c_n}.
\end{align}
\end{subequations}
Each one of the first four partial derivatives is not defined at two or four of the breakpoints $\{\bar{v}_n\pm\delta_n, \bar{v}_n\pm\sigma_n\}$. Landing on those points during the iterative process however is a zero-measure event. 


Evaluating gradient $\bg^i$ at $\bz^i$ requires computing the equilibrium injections $\{\bq_s(\bz^i)\}_{s=1}^S$ for $\bz^i$ at iteration $i$. These equilibrium injections can be computed either as the minimizer of \eqref{eq:inner}, or by simulating the grid dynamics in \eqref{eq:dynamics1} until convergence. The dynamics are guaranteed to converge as $\bz^i\in\mcZ_{\epsilon}$, and thus, yield stable rules for all PGD iterations $i$. Nonetheless, the settling time for the dynamics varies with $\epsilon$. The algorithm is summarized as Alg.~\ref{alg:PGD}. 

\begin{algorithm}[t]
	\caption{PGD for Optimal Volt/VAR Control Rule Design}\label{alg:PGD}
	\begin{algorithmic}[1]
		\renewcommand{\algorithmicrequire}{\textbf{Input:}}
		\renewcommand{\algorithmicensure}{\textbf{Output:}} 
		\REQUIRE Load/solar scenarios $\{(\bp_s^g,\bp_s^\ell,\bq_s^\ell)\}_{s=1}^S$
		\ENSURE Near-optimal Volt/VAR rule parameters $\bz$
		\STATE Compute $\tbv_s$ from \eqref{eq:vtilde} for $s=1:S$
		\STATE Find initial feasible $\bz^1$ by solving SOCP in \eqref{eq:project} for $\bx=\bzero$
		\WHILE{$\frac{|F(\bz^i)-F(\bz^{i-1}))|}{F(\bz^{i-1})} > 10^{-6}$}
		\STATE Solve the QP in \eqref{eq:inner} to find $\bq_s$ for $s=1:S$
		\STATE Compute $\bv_s$ from \eqref{eq:equilibrium_v} for $s=1:S$
		\STATE Compute gradient $\bg^i$ from \eqref{eq:dF}--\eqref{eq:derivatives}
        \STATE Update $\bz^{i+1}$ via \eqref{eq:pgd} by solving the SOCP in \eqref{eq:project}
        \ENDWHILE
	\end{algorithmic}
\end{algorithm}

\section{Numerical Tests}\label{sec:tests}

\begin{figure}[t!]
    \centering
    \includegraphics[width=0.9\linewidth]{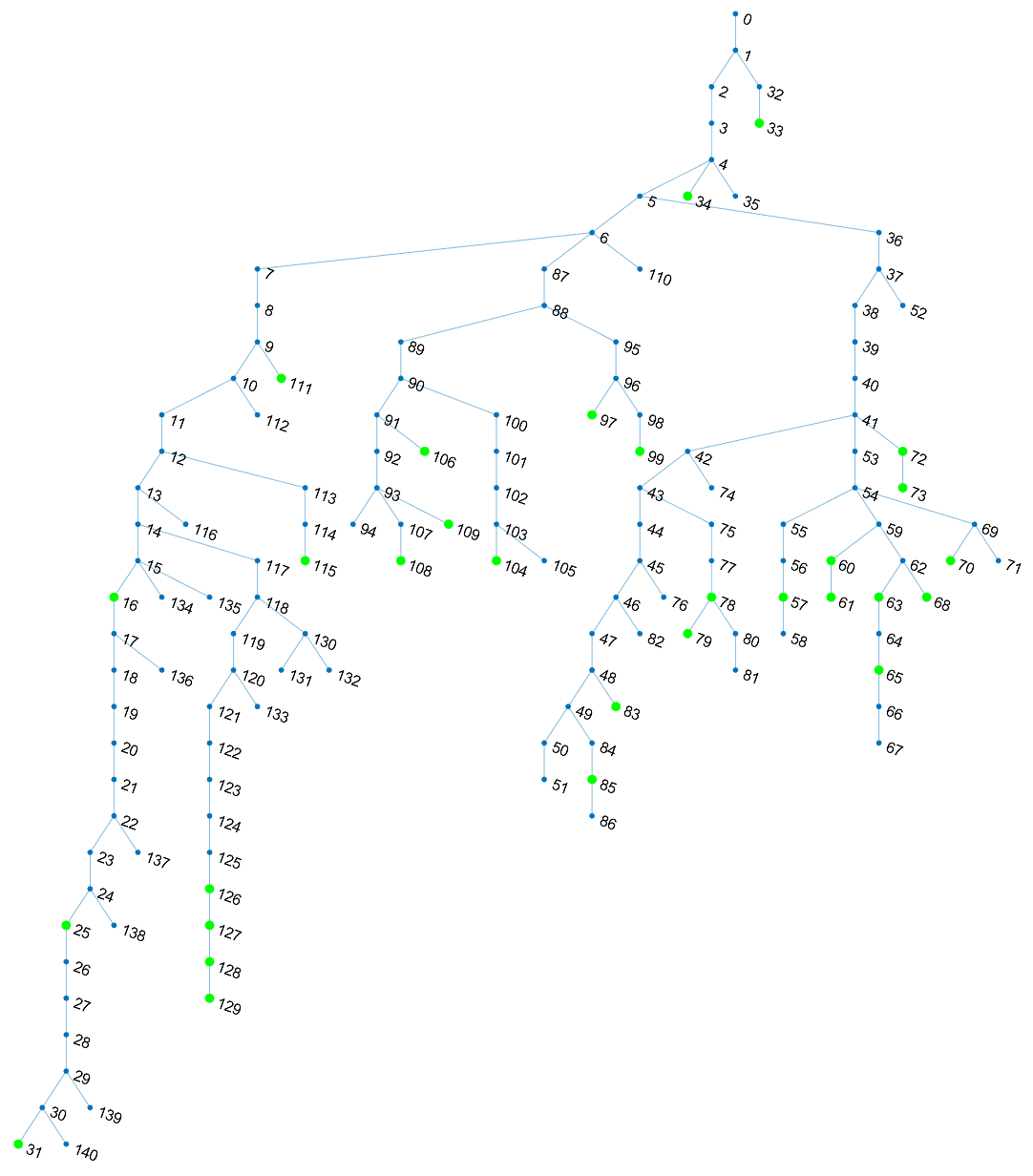}
    \caption{IEEE 141-bus feeder with 30 PVs added at highlighted green buses.}
    \label{fig:Case141PV30}
\end{figure}

The proposed Volt/VAR design was numerically evaluated using the IEEE 141-bus feeder converted to single-phase~\cite{KHODR20081192}. Since the original feeder had no solar generation units, we added 26 0.5MW PVs, and four 2MW PVs at buses 126, 127, 128, 129 as shown in Fig.~\ref{fig:Case141PV30}. All tests were run on an Intel Core i7–10700F CPU at 2.90 GHz, 32 GB RAM desktop computer with a 64-bit operating system. We used MATLAB 2022a and MATPOWER 7.1 for AC PF tests, and YALMIP and Gurobi 9.5.1; the code is available online at~\texttt{\url{https://github.com/IlgizMurzakhanov}}. 


Regarding data generation, real-world one-minute active load and solar generation data were extracted for April 2, 2011 from the Smart* project~\cite{Mishra2012Smart}. For active loads, homes with indices 20-369 were used. For each one of the 84 non-zero injection buses of the IEEE 141-bus system, the load was obtained upon averaging active loads of every 4 consecutive homes of the Smart* project. The values of active loads were subsequently scaled, so that maximum active load values per bus were 2.5 times the benchmark value. As the Smart* project does not provide reactive loads, reactive loads were synthetically generated from active loads using the power factors of the IEEE 141-bus benchmark. Active solar generation values were scaled, so the maximum value per PV matched its benchmark value. Volt/VAR rules were designed based on scenarios during the periods 09:00--11:00 and 13:30--15:30. For each of these two periods, the 120 one-minute data points were averaged per 5-minute intervals to produce a total of $S=24$ scenarios.

For our tests, the Volt/VAR parameters $\bz$ were designed using Algorithm~1. It is worth stressing the difference between the time steps of the dynamical system indexed by $t$ in \eqref{eq:dynamics1}, and the PGD iterations indexed by $i$ in \eqref{eq:pgd}. The former characterizes the number of time steps needed for the physical system to settle after changing $\tbv$; the latter determines the computational time needed for finding the optimal $\bz$. 

\begin{figure}[t]
    \centering
    \includegraphics[width=0.88\linewidth]{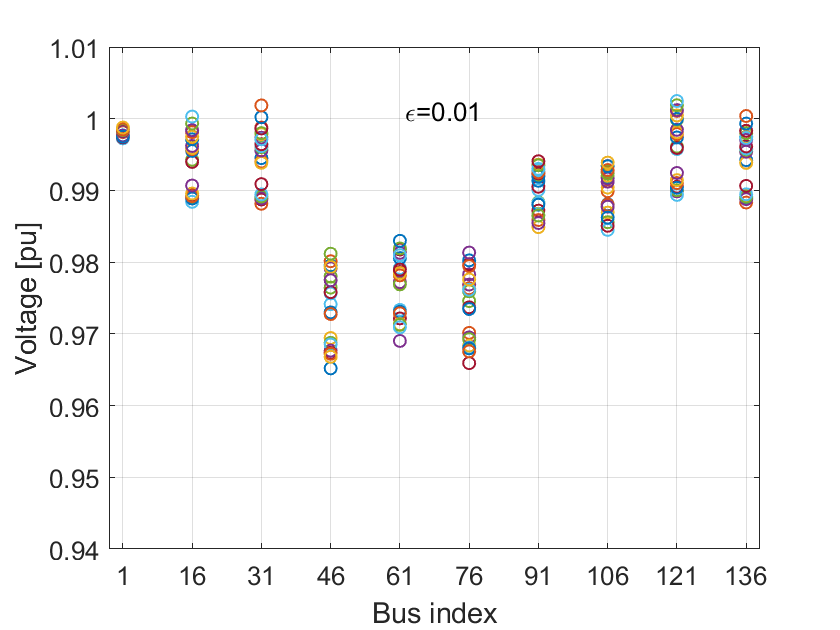}
    \includegraphics[width=0.88\linewidth]{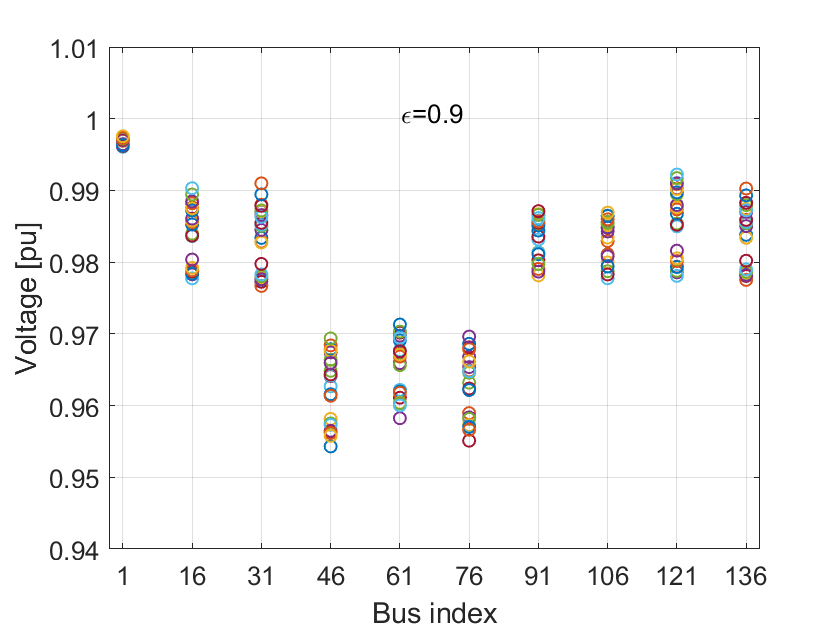}
    \includegraphics[width=0.88\linewidth]{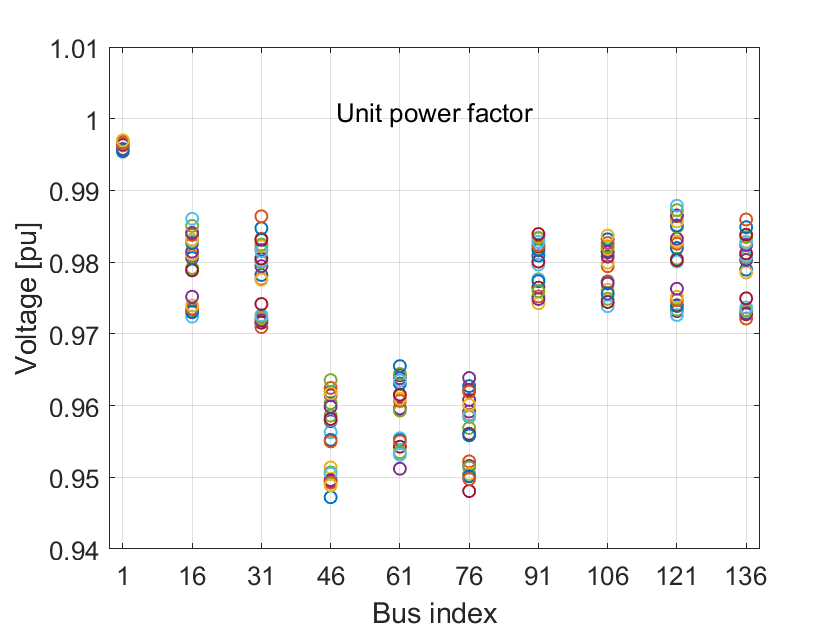}
    \caption{Voltage magnitudes during the 09:00--11:00 interval for optimized curve parameters using $\epsilon=0.01$ [top panel]; $\epsilon=0.9$ [middle panel]; and unit power factor (i.e. no reactive power support [bottom]. Voltages are shown on a subset of buses obtained upon sampling one every 15 buses. The different colors correspond to the 24 load and PV generation scenarios studied. }
    \label{fig:Voltages_10Buses_LPF_24sc_eps001_eps09_NAS}
\end{figure}

\begin{figure}[t]
    \centering
    \includegraphics[width=0.88\linewidth]{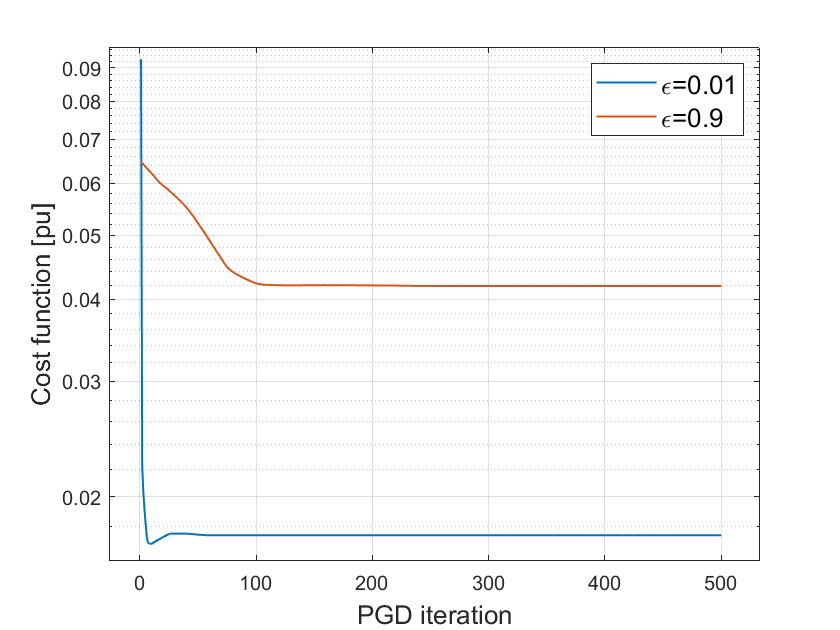}
    \vspace*{-1em}    
    \caption{Cost convergence of PGD iterations during the 9:00--11:00 interval. The option of $\epsilon = 0.01$ entails that the optimal curve parameters $\bz$ are selected from a larger feasible set compared to the option $\epsilon = 0.9$; and it can hence attain a smaller optimal voltage regulation cost.}
    \label{fig:cost_eps}
\end{figure}

\emph{Influence of Parameter $\epsilon$.} First, we experimented with the stability margin constant $\epsilon$. It is obvious from constraints \eqref{eq:set1:low} and \eqref{eq:soc:a} that larger values of $\epsilon$ yield a smaller feasible set $\mcZ_{\epsilon}$ for $\bz$. In other words, for larger $\epsilon$ the operator puts more emphasis on stability at the expense of attaining a possibly higher voltage regulation cost. This trade-off is supported by Fig.~\ref{fig:Voltages_10Buses_LPF_24sc_eps001_eps09_NAS} showing feeder voltages sampled every 15 buses computed by the linearized grid model across all $S=24$ scenarios. Parameter $\epsilon=0.01$ achieves a better voltage profile compared to the profile obtained with $\epsilon=0.9$. Either options attain improved voltage profiles compared to the profile shown at the bottom panel corresponding to inverters operating at unit power factor. The latter option of no reactive power support leads to voltage violations exceeding $5\%$ at various buses and different scenarios.

Fig.~\ref{fig:cost_eps} shows the convergence of the PGD iterations in terms of the cost function of \eqref{eq:bilevel}. Of course, smaller values of $\epsilon$ may lead to control rules that are closer to instability and experience longer settling times. In our tests, Volt/VAR dynamics settled in 8-9 time steps for $\epsilon=0.01$, and 2-3 steps for $\epsilon=0.9$. Considering the previous discussion, we fixed $\epsilon=0.01$ for the remaining tests.

\emph{Running Time.} The most time consuming steps of Alg.~\ref{alg:PGD} are: Step 4 where the QP of \eqref{eq:inner} is solved $S$ times per PGD iteration; and Step 7 where the SOCP of \eqref{eq:project} is solved once per PGD iteration. The QP took roughly $T_\text{QP}=0.0029$~sec per scenario, and the SOCP took approximately $T_\text{SOCP}=0.0593$~sec. All other steps had comparatively negligible times. The computation times for QP/SOCP were relatively invariant to $\epsilon$. Assuming no parallelism (the QPs can be solved in parallel), each PGD iteration took 0.13~sec. As PGD was terminated after about 500 iterations (cf. Fig.~\ref{fig:cost_eps}), it took roughly 1~minute.

\begin{figure}[t]
    \centering
    \includegraphics[width=0.88\linewidth]{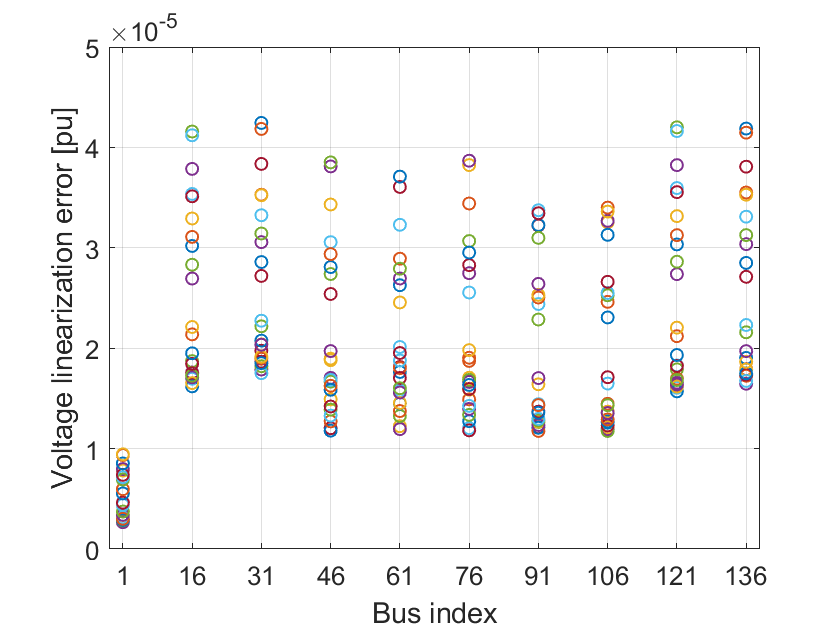}
    \caption{Error between voltage magnitudes at equilibrium obtained by the linearized and the AC grid model over $S=24$ scenarios during the 09:00--11:00 interval. Voltage are shown on a subset of buses obtained upon sampling one every 15 buses.}
    \label{fig:ACPF_LDF_10bus}
\end{figure}

\begin{figure}[t]
    \centering
    \includegraphics[width=0.88\linewidth]{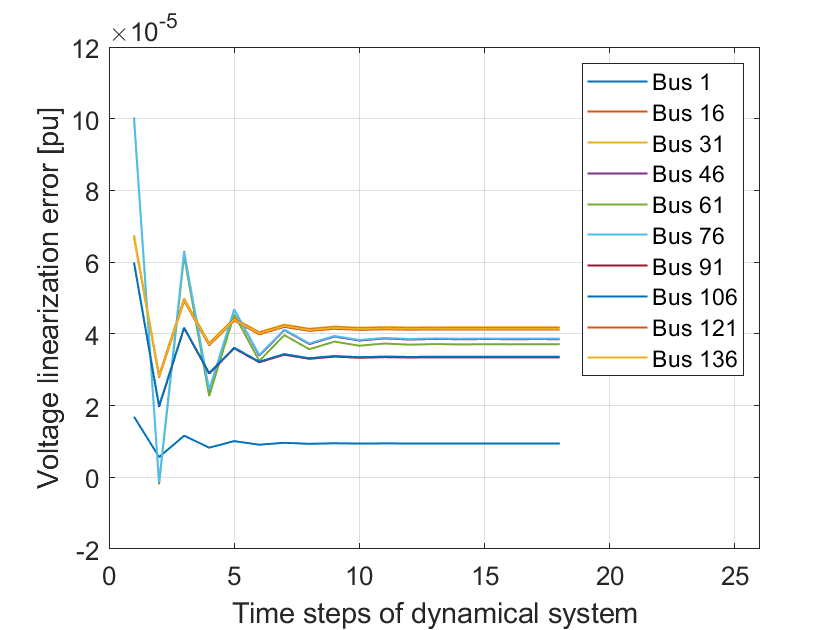}
    \caption{Error in dynamic voltages during Volt/VAR transients between the linearized and AC grid models over one scenario of the 9:00--11:00 period.}
    \label{fig:VoltError_LPF_ACPF_10bus}
\end{figure}

\emph{Linearized vs. AC Grid Model.} Thus far, the feeder has been modeled using the linearized model of \eqref{eq:ldf}. The next set of tests intends to evaluate the designed control rules using the actual AC grid model. To this end, we first found the optimal $\bz$ for the rules of Fig.~\ref{fig:curve} using the PGD algorithm. Then, the rules were applied iteratively on the AC grid model, i.e., equation \eqref{eq:dynamics1:a} was replaced by a power flow solver. {Fig.}~\ref{fig:ACPF_LDF_10bus} shows the error in voltages between the linearized and the AC grid model at equilibrium. As in Fig.~\ref{fig:Voltages_10Buses_LPF_24sc_eps001_eps09_NAS}, voltages are shown at a subset of 10 buses obtained by sampling one every 15 buses. As demonstrated by the plotted error, the linearized model provides a reasonable approximation with the largest deviation across buses and scenarios being less than $5 \cdot 10^{-5}$~pu. And this despite the fact that under the considered scenarios, the grid was heavily loaded and away from the unloaded conditions around which the linearized model has been derived. Beyond equilibrium voltages, an example of which we showed in Fig.~\ref{fig:Voltages_10Buses_LPF_24sc_eps001_eps09_NAS}, Volt/VAR dynamics also exhibited similar transient behavior under the linearized and the AC models: {Fig.}~\ref{fig:VoltError_LPF_ACPF_10bus} depicts the error in the dynamic evolution of voltages between the two models under one representative scenario. Similar behavior in terms of error and convergence within 5-10 steps was observed across scenarios. These two tests on equilibrium and dynamic voltages corroborate that the feeder can be well approximated by the linearized model. 

\begin{figure}[t]
    \centering
    \includegraphics[width=0.9\linewidth]{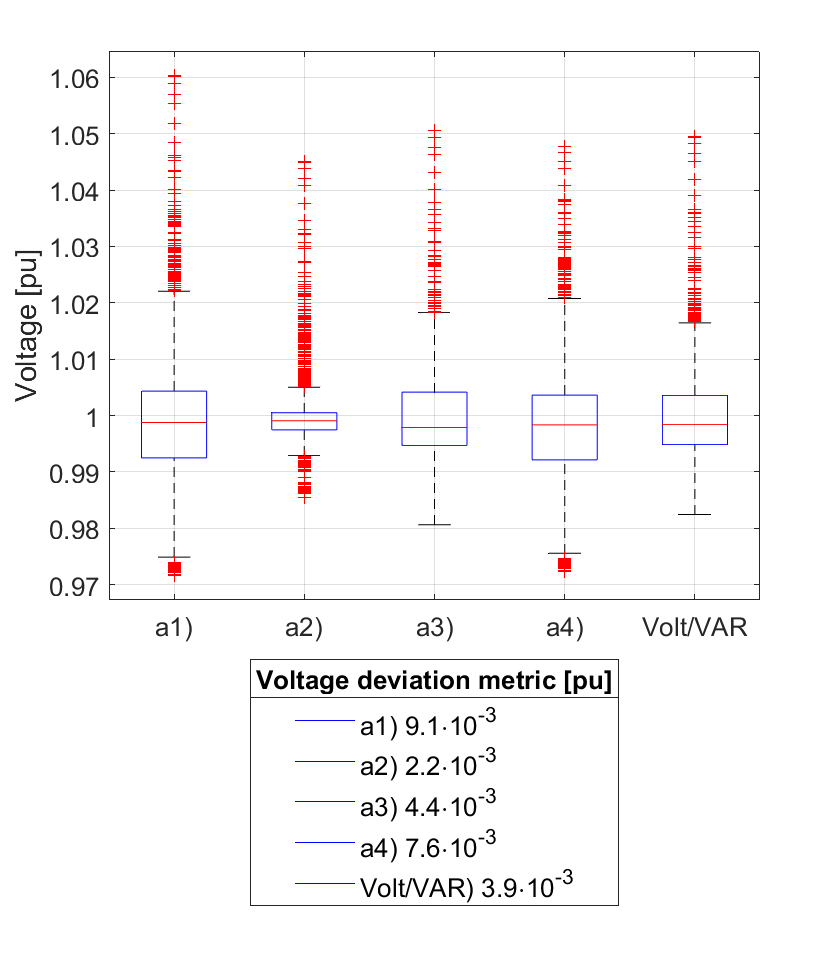}
    \vspace*{-2em}
    \caption{Boxplots of voltages during the 13:30--15:30 interval across all 140 buses and 24 scenarios using the four voltage regulation alternatives and the proposed customized Volt/VAR rules. Voltage regulation alternatives: a1) unit power factor operation, a2) per-scenario optimal operation, a3) optimal stochastic operation across scenarios, a4) default Volt/VAR settings.}
    \label{fig:alternatives}
\end{figure}

\emph{Comparison with Alternatives.} The proposed Volt/VAR approach is compared numerically to four inverter-based voltage regulation alternatives, which are detailed next.
\renewcommand{\theenumi}{\emph{a\arabic{enumi}}}
\begin{enumerate}
\item\emph{Unit power factor operation} under which inverters do not provide reactive power compensation.

\item\emph{Per-scenario optimal operation:} Inverter setpoints are optimized per scenario by solving the problems
\begin{align*}
\min_{-\hbq\leq \bq_s\leq \hbq}~\|\bX\bq_s + \tbv_s-\bone\|_2^2\quad \text{for}~s=1,\ldots,S.
\end{align*}
This scheme requires frequent two-way communication between the utility and DERs, but serves as a benchmark. 

\item\emph{Optimal stochastic operation across scenarios:} Inverter setpoints are updated once every two hours and they are determined as the minimizers of the stochastic problem
\begin{align*}
\min_{-\hbq\leq \bq\leq \hbq} ~\sum_{s=1}^S\|\bX\bq + \tbv_s-\bone\|_2^2.
\end{align*}
As with Volt/VAR rules, here the utility communicates with DERs only once over the 2-hr period to communicate the setpoint $\bq$. Different from the Volt/VAR rules however, setpoints here are not adaptive to grid conditions. 

\item\emph{Default Volt/VAR settings.} Here we implement the default settings of the IEEE 1547.8 Standard for DERs of Type B, which are recommended under high DER penetration with frequent large variations~\cite[Table~8]{IEEE1547.8}. 
\end{enumerate}

Apparently, scheme \emph{a2)} should attain lower \emph{voltage deviation metric} (VDM)
\[\text{VDM}:=\dfrac{1}{2S}\sum_{s=1}^S\|\bv_s-\bone\|_2^2\]
than scheme \emph{a3)}. The metric VDM is equivalent to the cost $F(\bz)$ the ORD task is trying to minimize over $\bz$. Our goal is for the optimized Volt/VAR curves to perform better than \emph{a3)} and \emph{a4)}, thus, hitting the sweet spot between communication and voltage regulation performance.

{Fig.}~\ref{fig:alternatives} depicts the distribution of voltages across all $N=140$ buses and $S=24$ scenarios using the four alternatives and the optimized Volt/VAR rules. Voltage profiles under \emph{a1)} exhibit unacceptable deviations. Scheme~\emph{a2)} offers acceptable voltage regulation at the expense of increased communication and computation overhead. Scheme \emph{a3)} violates the $5\%$ voltage limits. Scheme \emph{a4)} provides acceptable voltages, yet it attains higher VDM. It is should also be noted that for this particular setup, the default Volt/VAR settings did not satisfy the stability constraint \eqref{eq:stability2}, but did satisfy the stability constraint $\|\bA\bX\|_2<1-\epsilon$ relying on the spectral norm. This shows that the restriction of \eqref{eq:stability1} by \eqref{eq:stability2} is not tight as expected. Despite this particular setup, the default are not necessarily stable under all setups. For example, if thirty 1 MW-solar PV units are installed on buses 20, 122, 50, 84, 67, and downwards towards the leaves, they do violate the stability condition of \eqref{eq:stability1} and consequently \eqref{eq:stability2}. The customized Volt/VAR curves provide practically relevant solutions, offering more concentrated voltage profiles than \emph{a3)}-\emph{a4)} due to their scenario-adaptive nature. The same figure also reports the VDM for all four schemes, and corroborates the superiority of Volt/VAR curves over \emph{a3)} and \emph{a4)}. 

\begin{figure}[t]
    \centering
    \includegraphics[width=0.9\linewidth]{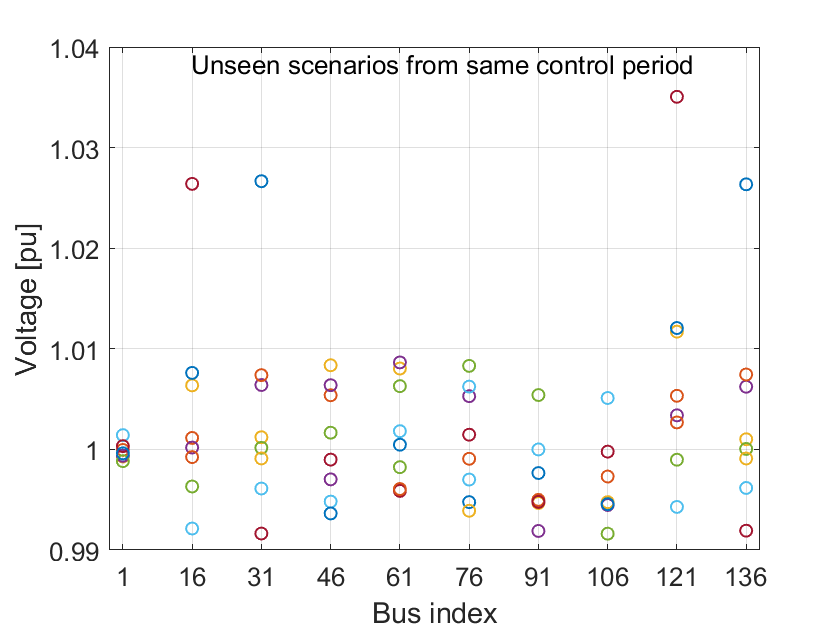}
    \includegraphics[width=0.9\linewidth]{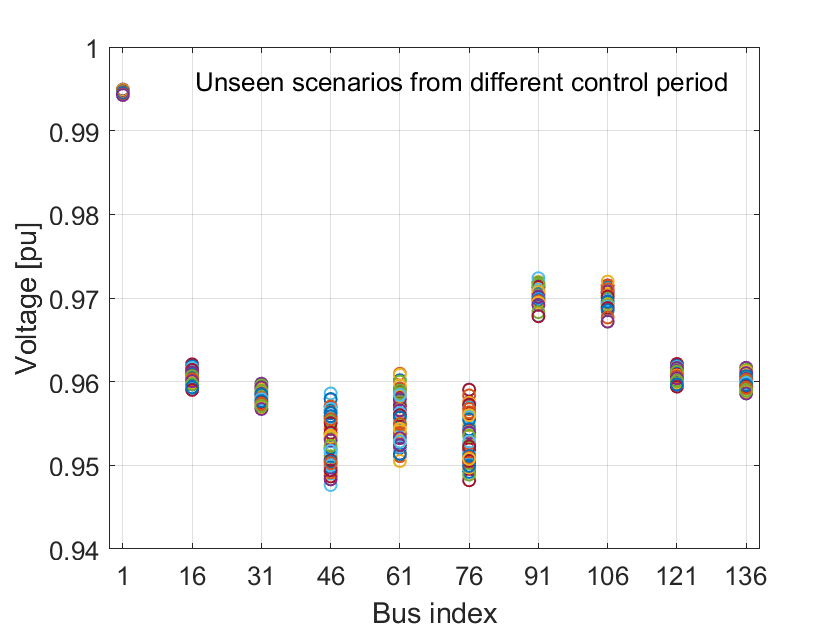}
    \caption{Voltages experienced under \emph{unseen} or \emph{out-of-sample} loading scenarios, that is scenarios not used while designing the Volt/VAR control curves in \eqref{eq:outer}. Rules here were trained using scenarios drawn from the 13:30--15:30 period. Unseen scenarios were drawn from the same period \emph{(top)}, and the period 19:00-20:00 \emph{(bottom)}.}
    \label{fig:VOutOfSample}
\end{figure}

\emph{Out-of-Sample Loading Scenarios.} How do the designed rules perform under \emph{unseen} or \emph{out-of-sample} loading scenarios, that is scenarios not used while solving \eqref{eq:outer}? To answer this question, we designed the rules using a subset of scenarios, and evaluated their voltage regulation performance on a different subset of scenarios. We conducted two tests. In the first test, we designed rules using $S=16$ scenarios drawn from the 13:30--15:30 control period, and evaluated voltage profiles under another sample of 8 scenarios from the same control period; see top panel of Fig.~\ref{fig:VOutOfSample}. In the second test, rules were trained again using scenarios from the 13:30--15:30 control period, but tested on the 19:00--20:00 control period; see bottom panel of Fig.~\ref{fig:VOutOfSample}. The tests show that the designed rules perform well under same-period scenarios, where voltages are maintained within $\pm 5\%$. On the contrary, the rules perform poorly under scenarios drawn from another control period. This proves the necessity for redesigning the control rules periodically through the day to adjust to different loading conditions. 

\begin{figure}[t]
    \centering
    \includegraphics[width=0.9\linewidth]{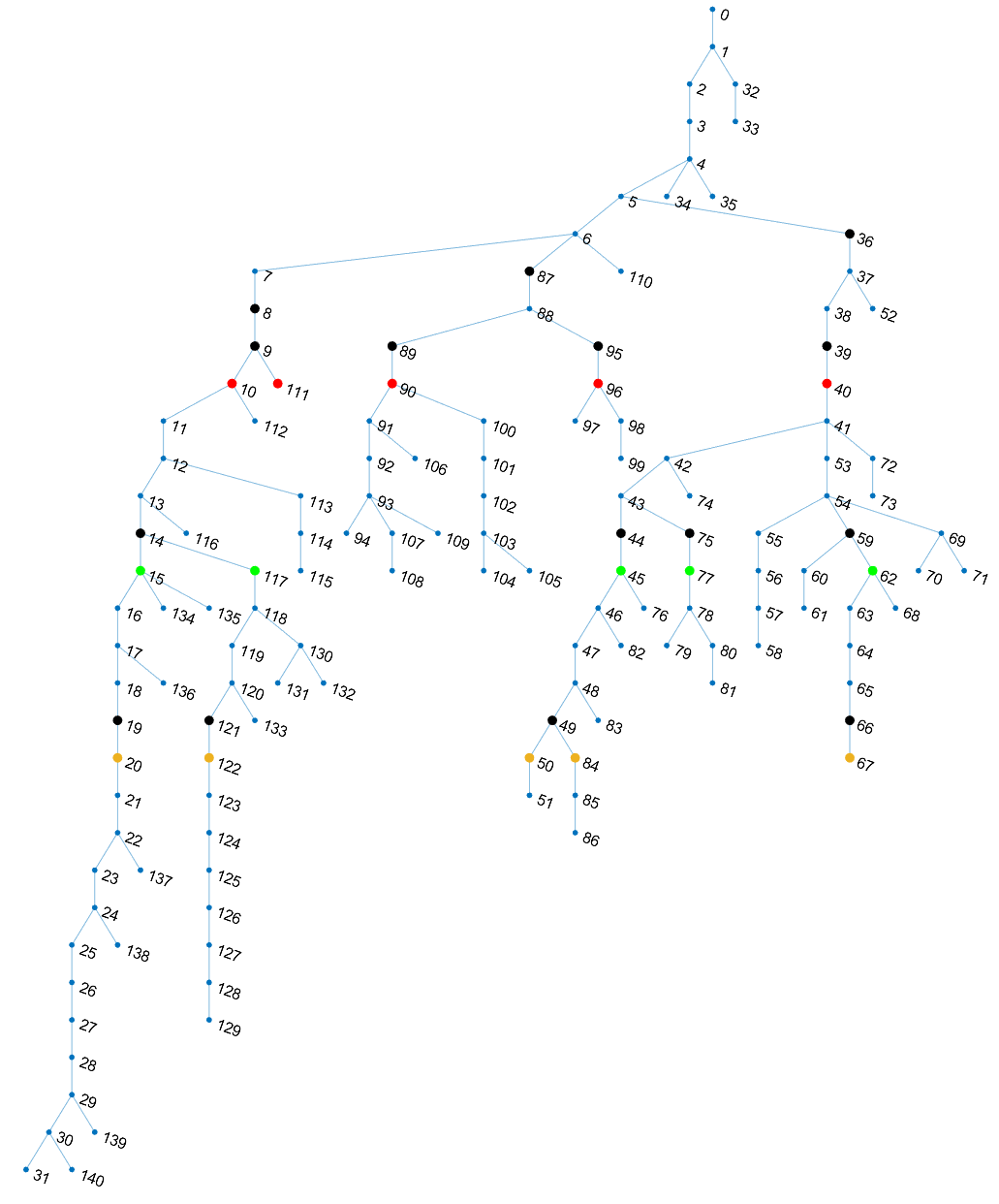}
    \caption{Tested IEEE 141-bus benchmark feeder with added 5 PV units at depth-10 (highlighted red), depth-15 (highlighted green), depth-20 (highlighted orange), and 15 PVs continuously operating at unit power factor (highlighted black).}
    \label{fig:VarDepths}
\end{figure}

\begin{figure}[t]
    \centering
    \includegraphics[width=0.9\linewidth]{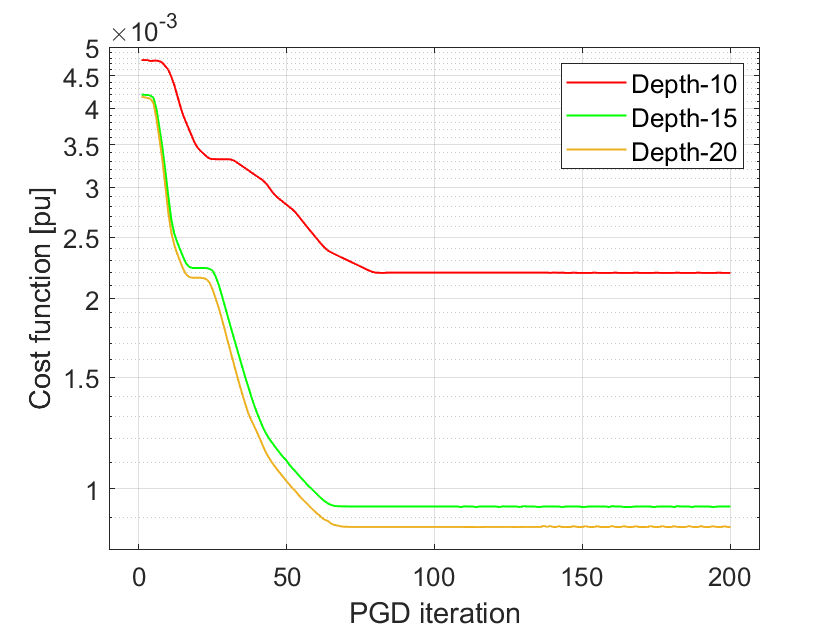}
    \caption{Convergence in terms of the cost function \eqref{eq:bilevel} for the 13:30--15:30 control interval under $\epsilon = 0.01$ after placing 5 upgraded inverters at different depths. Placing smart inverters further away from the substation seems to be offering better voltage regulation performance (lower VDM).}
    \label{fig:CompCostFunc_VarDepths}
\end{figure}

\begin{figure}[t]
    \centering
    \includegraphics[width=0.7\linewidth]{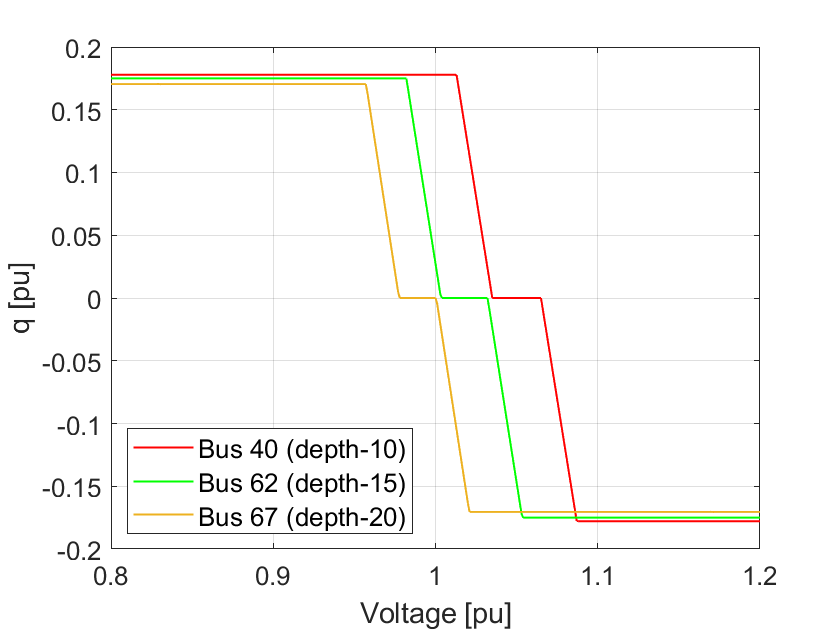}
    \caption{{Designed Volt/VAR rules for the 13:30--15:30 interval under $\epsilon = 0.01$ for 3 buses with upgraded inverters, placed at different depths. 
    }}
    \label{fig:VoltVar_3VarDepths}
\end{figure}

\begin{figure}[t]
    \centering
    \includegraphics[width=0.7\linewidth]{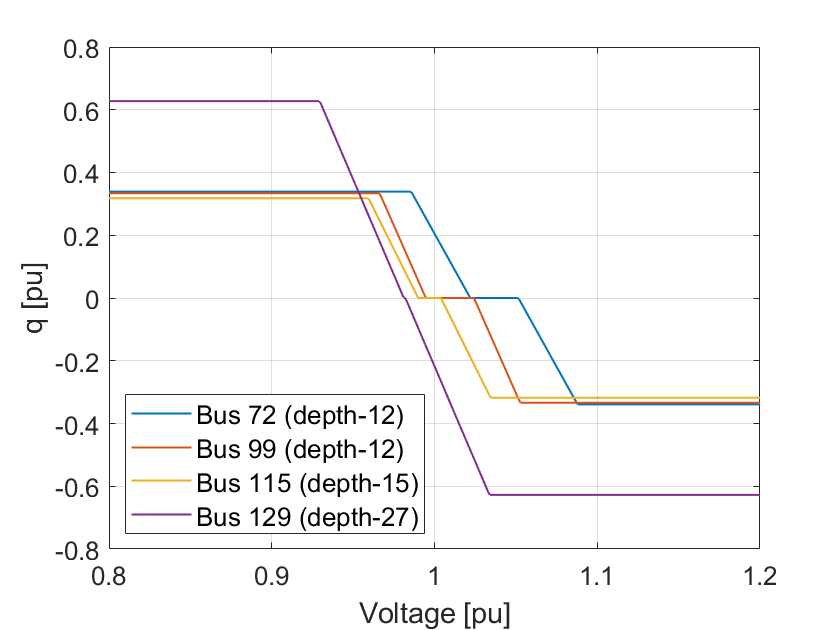}
    \caption{{Designed Volt/VAR rules for the 13:30--15:30 interval under $\epsilon = 0.01$ for 4 inverters in the IEEE 141-bus feeder with 30 PVs.}}
    \label{fig:VoltVar_30PVs}
\end{figure}

\color{black}
\emph{Effect of Inverter Positions.} This last test aimed at evaluating the effect of Volt/VAR control by inverters placed at different locations. The motivation is that an operator may want to have most inverters operating at unit power factor and only a handful of them being upgraded to support the functionality of Volt/VAR control. The goal of this test is not to optimally place smart inverters, but rather, to infer some qualitative conclusions on how the location and the number of inverters affects the performance and possibly the shape of optimal Volt/VAR curves. The conjecture is that inverters located at buses further away from the substation could have higher impact on voltage regulation. This is because the columns of sensitivity matrix $\bX$ in \eqref{eq:ldf} have larger values for buses further away from the substation.  Nonetheless, for the exact same buses, slopes are most limited by stability constraint \eqref{eq:stability2:C2}. In particular, the curve slope $\alpha_n$ at bus $n$ is upper bounded at least by the inverse of the sum $\sum_{m\in\mcG}X_{nm}$ as dictated by \eqref{eq:stability2:C2}. To explore this issue, we conducted tests where again 30 inverters have been installed on the IEEE 141-bus feeder as before, but now only 5 of them provide the Volt/VAR functionality and the remaining 25 operate at unit power factor. Note that these 25 PVs include 15 PVs which continuously operate at unit power factor, and two sets of 5 PVs which operate at unit power factor only within the current test and provide Volt/VAR functionality when selected. We tested three setups, where the 5 smart inverters were placed at depths of 10, 15, and 20. Placing an inverter at depth 10 means that it is sited at a bus that is 10 buses away from the substation; see Fig.~\ref{fig:VarDepths}. 

Tests illustrated on Fig.~\ref{fig:CompCostFunc_VarDepths} suggest that upgrading first inverters further away from the substation is preferable as it provides better voltage profile. Note that the runtime per PGD iteration in Fig.~\ref{fig:CompCostFunc_VarDepths} is comparable to the earlier system with 30 upgraded PVs (0.13 seconds). {Fig.}~\ref{fig:VoltVar_3VarDepths} shows the optimal curves for three inverters sited at different depths. Several conclusions can be made. 

First, due to local generation and reduced local load, buses at greater depths experience higher voltages. As a result, the optimal $\bar{v}$ for those Volt/VAR-enabled inverters is further to the left compared to inverters closer to the substation in Fig.~\ref{fig:VoltVar_3VarDepths}. An additional interesting observation about the optimal $\bar{v}$ is related to inverters at the same depth, such as at buses 72 and 99 shown in Fig.~\ref{fig:VoltVar_30PVs}. While buses 72 and 99 are at the same depth, they have different $\bar{v}$ values. This is related to the fact that bus 99 has no load, while bus 72 has a load. As a result, bus 99 experiences higher voltage, and its $\bar{v}$ shifts more to the left compared to bus 72. Second, inverters at greater depth are slightly further away from saturating $\hat{q}$ limits, which could be attributed to the tighter stability constraints as well as the fact that buses at larger depths correspond to larger entries of $\bX$. Third, inverters at smaller depths have wider deadband (i.e. larger $\delta$). Compare for example the inverter in bus 40 (depth-10) versus bus 67 (depth-20) in Fig.~\ref{fig:VoltVar_3VarDepths}, or the inverters on buses 72 and 99 (depth-12) vs bus 129 (depth-27) in Fig.~\ref{fig:VoltVar_30PVs}; $\delta$ for bus 129 is almost 0. 

Inverters at smaller depths have wider deadband for two reasons. First, they wait for the inverters at larger depths to react first, as this is more effective for the voltage regulation. This confirms our conjecture that inverters located at buses further away from the substation have higher impact on voltage regulation. {A similar observation was made in \cite{Baker18}, which further strengthens the proposed premise.} Second, by waiting to react a bit later, they can cover a wider range of voltages. This is clearer in Fig.~\ref{fig:VoltVar_3VarDepths}, where all inverters have the same nominal capacity, and we observe they have the same slope. Indeed, in the case where only 5 inverters are equipped with Volt/VAR functionality, the optimizer requires $\sigma - \delta$ to be as small as possible, and $\hat{q}$ to be as high as possible in order to regulate voltage effectively. This means that for all 5 inverters \eqref{eq:lb_sigmadelta} is binding at the lower bound. In such a case, a larger $\delta$ and a similar $\sigma - \delta$  means larger $\sigma$, i.e. wider voltage range. As a result, inverters at smaller depths have a wider voltage range. In case all 30 inverters were equipped with Volt/VAR functionality, shown in Fig.~\ref{fig:VoltVar_30PVs}, we not only observe that the slopes are no longer the same, but also not all inverters set $\hat{q}$ close to their nominal capacity. This happens because in such a case there are enough control resources available, so the optimizer does not need to ``max-out'' the settings of the available controls as in the case with 5 inverters. Still, we continue to observe that inverters at smaller depths, e.g. at buses 72 and 99, have wider voltage ranges than inverters at larger depths, such as at buses 115 and 129.

\section{Conclusions}\label{sec:conclusions}
We have proposed a novel methodology for designing Volt/VAR control rules. Different from existing alternatives, the designed rules are compliant with the IEEE 1547 and ensure grid dynamic stability. Using the proposed PGD-based algorithm, utilities can adjust Volt/VAR controls per bus and predicted grid scenarios. Our numerical tests have corroborated that: \emph{i)} the designed rules can effectively respond to varying conditions; \emph{ii)} linearized dynamics behave sufficiently close to AC dynamics; \emph{iii)} the margin $\epsilon$ relates to settling times and optimality; \emph{iv)} the designed rules perform better than an one-size-fits-all inverter dispatch and worse than per-scenario optimal dispatches; \emph{v)} PGD iterations can be completed within minutes thus offering a lucrative solution; and \emph{vi)} upgrading inverters towards the feeder ends seems to be more effective. 

\bibliographystyle{IEEEtran}
\bibliography{myabrv,power,kekatos,inverters}

\begin{IEEEbiography}[{\includegraphics[width=1in,height=1.25in,clip,keepaspectratio]{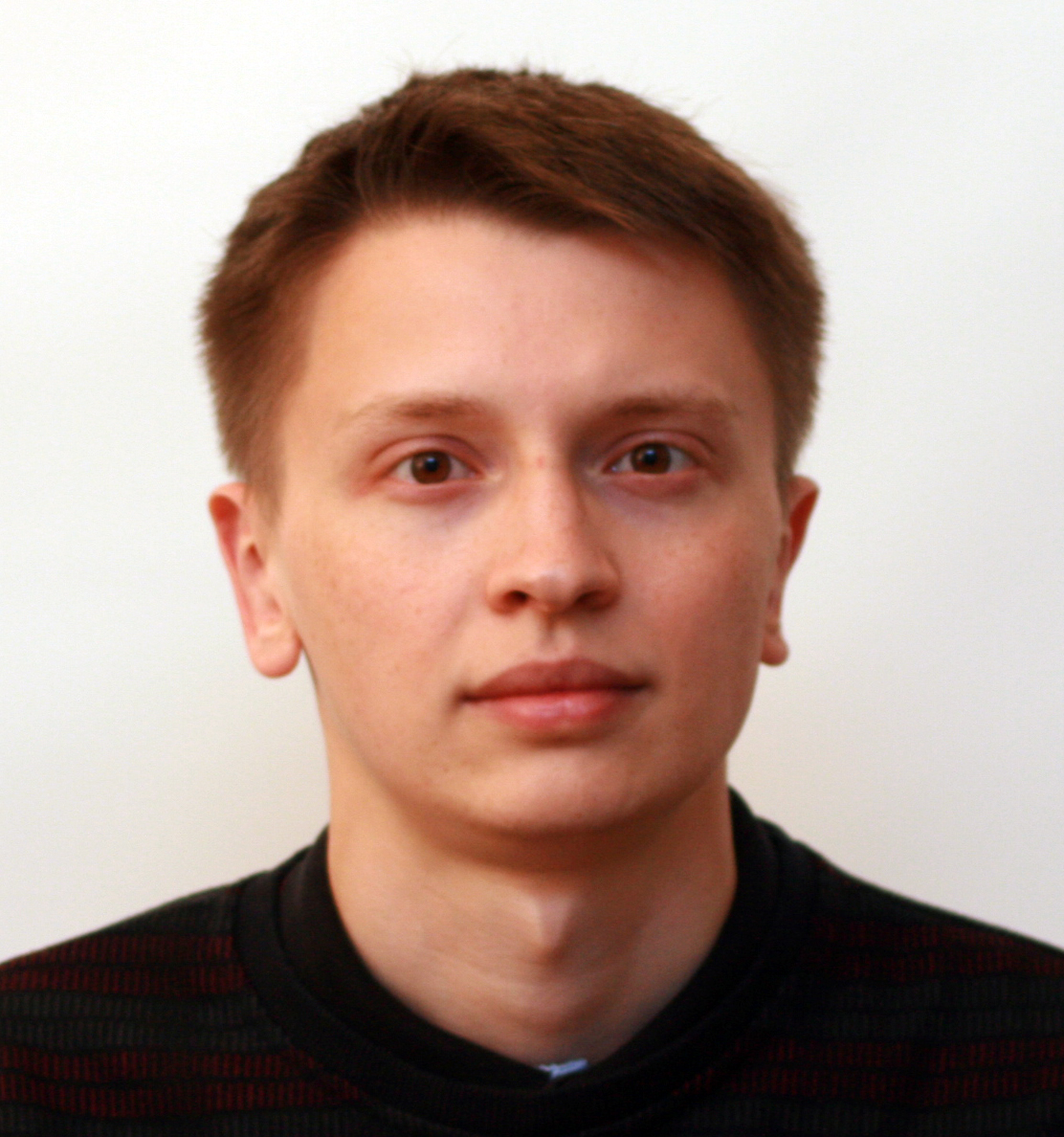}}] {Ilgiz Murzakhanov} (M'19) received the B.Eng. degree in electric power engineering and electrical engineering from the Moscow Power Engineering Institute, Russia, in 2016, the M.Sc. degree in energy systems from the Skolkovo Institute of Science and Technology, Russia, in 2018, and the Ph.D. degree in electrical and computer engineering from the Technical University of Denmark (DTU), in 2023. In 2017 and 2022, he was a Visiting Researcher at the Delft University of Technology, Netherlands, and Virginia Tech, USA, respectively. He is a Postdoctoral Researcher with the Department of Wind and Energy Systems, DTU. He is currently working on physics-informed neural networks and neural network verification.
\end{IEEEbiography}

\begin{IEEEbiography}[{\includegraphics[width=1in,height=1.25in,clip,keepaspectratio]{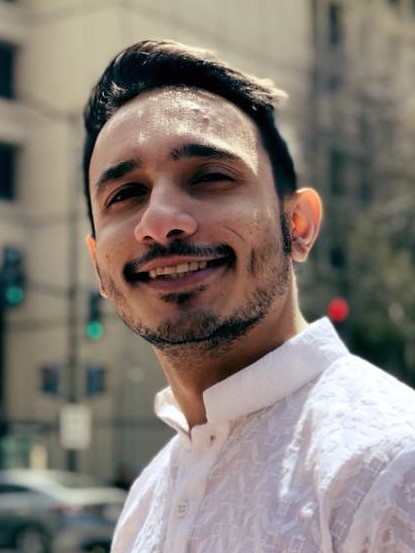}}] {Sarthak Gupta} received the B.Tech. degree in Electronics and Electrical Engineering from the Indian Institute of Technology Guwahati, India, in 2013; and the M.Sc. and Ph.D. degrees in Electrical Engineering from Virginia Tech, Blacksburg, VA, USA, in 2017 and 2022, respectively. During 2017-2019, he worked as an Associate Engineer with the Distributed Resources Operations Team of the New York Independent System Operator, NY, USA. During the summer of 2021, he interned at the Applied Mathematics and Plasma Physics Group of the Los Alamos National Laboratory, Los Alamos, NM, USA. He is currently a Senior Data Scientist with C3.AI. His research interests include deep learning, reinforcement learning, optimization, and power systems.
\end{IEEEbiography}

\begin{IEEEbiography}[{\includegraphics[width=1in,height=1.25in,clip,keepaspectratio]{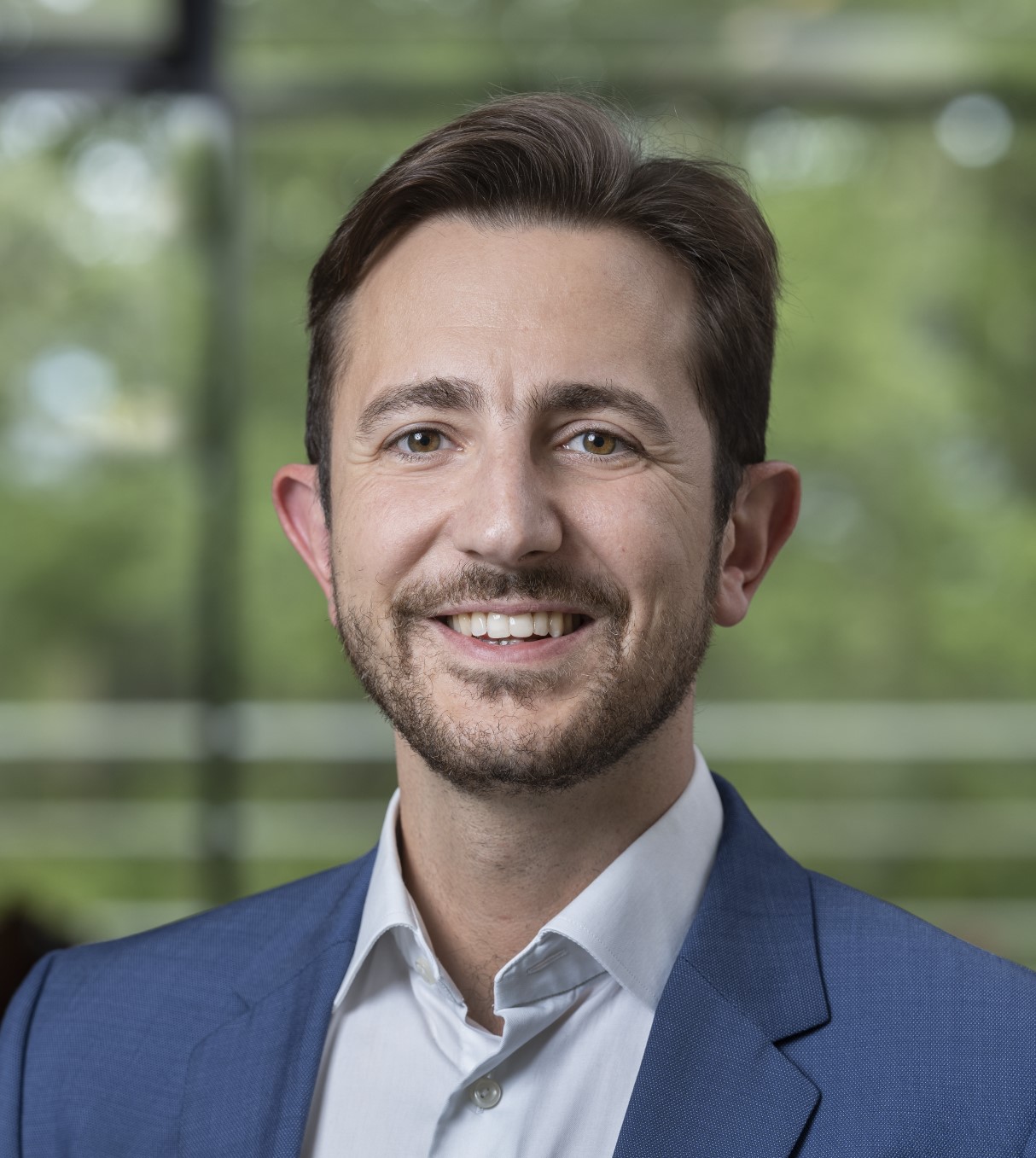}}] {Spyros Chatzivasileiadis} (S’04, M’14, SM’18) is the Head of Section for Power Systems and an Associate Professor at the Technical University of Denmark (DTU). Before that he was a postdoctoral researcher at the Massachusetts Institute of Technology (MIT), USA and at Lawrence Berkeley National Laboratory, USA. Spyros holds a PhD from ETH Zurich, Switzerland (2013) and a Diploma in Electrical and Computer Engineering from the National Technical University of Athens (NTUA), Greece (2007). He is currently working on trustworthy machine learning for power systems, quantum computing, and on power system optimization, dynamics, and control of AC and HVDC grids. Spyros is the recipient of an ERC Starting Grant in 2020, and has received the Best Teacher of the Semester Award at DTU Electrical Engineering.
\end{IEEEbiography}

\begin{IEEEbiography}[{\includegraphics[width=1in,height=1.25in,clip,keepaspectratio]{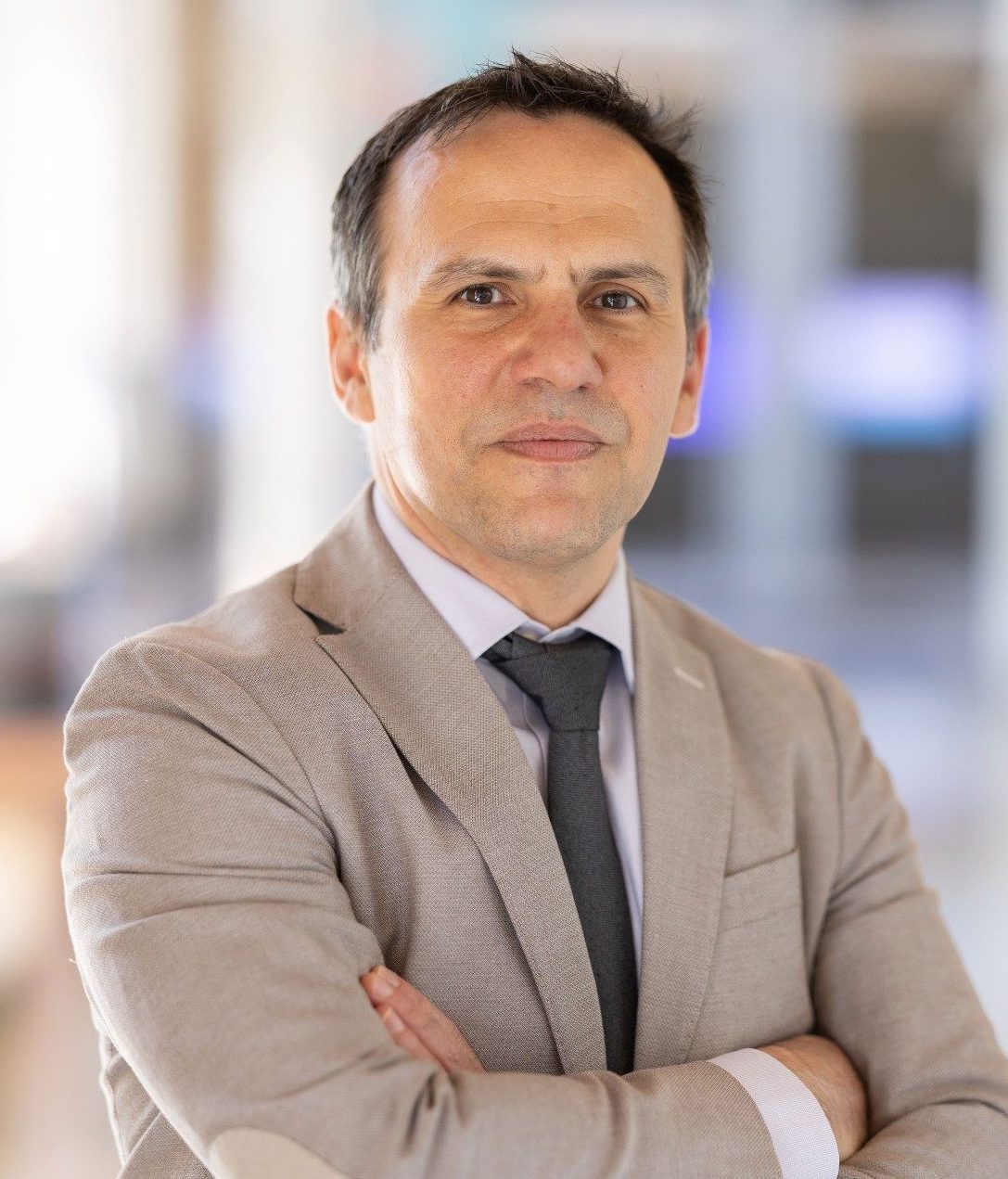}}] {Vassilis Kekatos} (SM'16) is an Associate Professor with the Bradley Dept. of ECE at Virginia Tech. He obtained his Diploma, M.Sc., and Ph.D. in computer science and engineering from the Univ. of Patras, Greece, in 2001, 2003, and 2007, respectively. He is a recipient of the NSF Career Award in 2018 and the Marie Curie Fellowship. He has been a research associate with the ECE Dept. at the Univ. of Minnesota, where he received the postdoctoral career development award (honorable mention). During 2014, he stayed with the Univ. of Texas at Austin and the Ohio State Univ. as a visiting researcher. His research focus is on optimization and learning for smart energy systems. During 2015-2022, he served on the editorial board of the IEEE Trans. on Smart Grid.
\end{IEEEbiography}
\end{document}